\documentclass{vldb}
\usepackage{thm-restate}
\usepackage{graphicx}
\usepackage{hyperref}
\usepackage{tikz}
\usepackage{adjustbox}
\usepackage{ntheorem}
\usepackage{amsmath}
\usepackage{makecell}
\usepackage[normalem]{ulem}
\usepackage{floatrow}
\usepackage{makecell}

\newfloatcommand{capbtabbox}{table}[][\FBwidth]
\usepackage{blindtext}
\usepackage[normalem]{ulem}
\usepackage{times}
\usepackage[show]{chato-notes}
\usepackage{cite}
\usepackage{color}
\usepackage{graphics}
\usepackage{graphicx}
\usepackage{mathtools}
\usepackage{xspace}
\usepackage{algorithm}

\usepackage{epsfig}
\usepackage{epstopdf}
\usepackage{booktabs}
\usepackage[font={bf,small}]{caption}
\DeclareCaptionType{copyrightbox}
\usepackage{xcolor}
\usepackage{balance}
\usepackage{enumitem}
\usepackage{framed}
\usepackage{multirow}
\usepackage{algpseudocode}
\usepackage{hyperref}
\hypersetup{
	colorlinks,breaklinks,
	urlcolor=[rgb]{0,0.25,0.75},
	linkcolor=[rgb]{0,0.25,0.75},
	citecolor=[rgb]{0,0.25,0.75}
}

\vldbTitle{}
\vldbAuthors{}
\vldbDOI{}
\vldbVolume{}
\vldbNumber{}
\vldbYear{}

\DeclareMathOperator*{\argmax}{arg\,max}
\DeclareMathAlphabet\mathbfcal{OMS}{cmsy}{b}{n}

\newcommand{\weic}[1]{{\color{blue} [\text{Wei Chen: }  #1]}}

\newcommand{\red}[1]{\textcolor{red}{#1}}

\newcommand{\pink}[1]{\textcolor{black}{#1}}

\newcommand{\spara}[1]{\vspace{1mm}\noindent\textbf{#1.}}

\newcommand{\Added}{{\it Added}}

\newcommand{\eat}[1]{}

\newcommand{\InFullOnly}[1]{}
 
\theoremstyle{plain}

\newtheorem{theorem}{Theorem}
\newtheorem*{theorem-non}{Theorem}

\newtheorem{claim}{Claim}

\newtheorem{definition}{Definition}

\newtheorem{lemma}{Lemma}

\newtheorem{problem}{Problem}

\newtheorem{proposition}{Proposition}
\newtheorem{sub-proposition}{Sub-proposition}

\def\noflash#1{\setbox0=\hbox{#1}\hbox to 1\wd0{\hfill}}

\mathchardef\mhyphen="2D %define a hyphen used in the math mode

\newcommand{\E}{{\mathbb{E}}\xspace}

\newcommand{\gm}{{\rm greedyWM}\xspace}  
\newcommand{\tc}{{\rm TCIM}\xspace}
\newcommand{\bc}{{\rm Balance{\text -}C}\xspace}

\newcommand{\mgrd}{{\rm MaxGRD}\xspace}
\newcommand{\sgrd}{{\rm SeqGRD}\xspace}
\newcommand{\supgrd}{{\rm SupGRD}\xspace}
\newcommand{\msgrd}{{\rm MaxSeqGRD}\xspace}

\newcommand{\mgrdp}{{\rm MaxGRDProc}\xspace}
\newcommand{\sgrdp}{{\rm SeqGRDProc}\xspace}
\newcommand{\sgrdi}{{\rm SeqGRD{\text -}NM}\xspace}

\newcommand{\umin}{{u_{\rm min}} \xspace}
\newcommand{\umax}{{u_{\rm max}} \xspace}

\newcommand{\util}{\mathcal{U}\xspace}
\newcommand{\utilt}{\mathcal{U}^+\xspace}

\newcommand{\ua}{{u}\xspace}

\newcommand{\price}{\mathcal{P}\xspace}
\newcommand{\val}{\mathcal{V}\xspace}
\newcommand{\noise}{\mathcal{N}\xspace}

\newcommand{\squishlist}{
 \begin{list}{$\bullet$}
  {  \setlength{\itemsep}{0pt}
     \setlength{\parsep}{3pt}
     \setlength{\topsep}{3pt}
     \setlength{\partopsep}{0pt}
     \setlength{\leftmargin}{2em}
     \setlength{\labelwidth}{1.5em}
     \setlength{\labelsep}{0.5em}
} }
\newcommand{\squishlisttight}{
 \begin{list}{$\bullet$}
  { \setlength{\itemsep}{0pt}
    \setlength{\parsep}{0pt}
    \setlength{\topsep}{0pt}
    \setlength{\partopsep}{0pt}
    \setlength{\leftmargin}{2em}
    \setlength{\labelwidth}{1.5em}
    \setlength{\labelsep}{0.5em}
} }

\newcommand{\squishdesc}{
 \begin{list}{}
  {  \setlength{\itemsep}{0pt}
     \setlength{\parsep}{2pt}
     \setlength{\topsep}{2pt}
     \setlength{\partopsep}{0pt}
     \setlength{\leftmargin}{2em}
     \setlength{\labelwidth}{1.5em}
     \setlength{\labelsep}{0.5em}
} }

\newcommand{\squishdesctight}{
 \begin{list}{}
  {  \setlength{\itemsep}{0pt}
     \setlength{\parsep}{0pt}
     \setlength{\topsep}{0pt}
     \setlength{\partopsep}{0pt}
     \setlength{\leftmargin}{1em}
     \setlength{\labelwidth}{1.5em}
     \setlength{\labelsep}{0.5em}
} }

\newcommand{\squishnumlist} {
\newcounter{qcounter}
\begin{list}{\arabic{qcounter}.~}{\usecounter{qcounter}} 
{  \setlength{\itemsep}{0pt}
    \setlength{\parsep}{0pt}
    \setlength{\topsep}{0pt}
    \setlength{\partopsep}{0pt}
    \setlength{\leftmargin}{2em}
    \setlength{\labelwidth}{1.5em}
    \setlength{\labelsep}{0.5em}
}}

\newcommand{\squishend}{
  \end{list}
}

\newcommand{\user}{{u}\xspace}

\newcommand{\itemset}{{I}\xspace}

\newcommand{\parameterset}{{\sf Param}\xspace}
\newcommand{\comic}{Com-IC\xspace}
\newcommand{\model}{UIC\xspace}
\newcommand{\prob}{CWelMax\xspace}

\newcommand{\allitems}{\mathbf{I}}
\newcommand{\allalloc}{\mathbfcal{S}}
\newcommand{\allallocp}{\mathbfcal{S}^P}

\newcommand{\alliseeds}{S_i}

\newcommand{\1}{{i_1}\xspace}
\newcommand{\2}{{i_2}\xspace}
\newcommand{\3}{{i_3}\xspace}
\newcommand{\4}{{i_4}\xspace}

\newcommand{\cA}{\mathcal{A}}

\newcommand{\cU}{\mathcal{U}}

\newcommand{\greedSeeds}{S^{\it Grd}}

\newcommand{\greedAlloc}{\mathbfcal{S}^{\it Grd}}

\newcommand{\seqAlloc}{\mathbfcal{S}^{\it Seq}}
\newcommand{\seqSeeds}{S^{\it Seq}}
\newcommand{\maxAlloc}{\mathbfcal{S}^{\it Max}}
\newcommand{\maxSeeds}{S^{\it Max}}

\newcommand{\greedSeq}{\mathbfcal{S}^{\it Seq}}

\newcommand{\optAlloc}{\mathbfcal{S}^{\it OPT}}

\newcommand{\sw}{\rho}

\newcommand{\OPT}{{\it OPT}}

\newcommand{\aware}{\mathcal{R}}
\newcommand{\awares}{\mathcal{R}^{\allalloc}}
\newcommand{\awarews}{\mathcal{D}_{w}^{\allalloc}}
\newcommand{\adopt}{\mathcal{A}}
\newcommand{\adopts}{\mathcal{A}^{\allalloc}}
\newcommand{\adoptws}{\mathcal{A}_{w}^{\allalloc}}

\newcommand{\budgetSwitch}{{\it budgetSwitch}}
\newcommand{\true}{{\bf true}}
\newcommand{\false}{{\bf false}}

\newcommand{\bvec}{\vec{b}} 
\newcommand{\bmax}{\overline{b}}

\newcommand{\calA}{\mathcal{A}}

\newcommand{\calJ}{\mathcal{J}}

\newcommand{\inst}{\mathcal{I}}
\newcommand{\calf}{\mathcal{F}}
\newcommand{\ground}{X}
\newcommand{\cover}{\mathcal{C}}

\newcommand{\PRIMM}{\textsf{PRIMA}}
\newcommand{\PRIMAP}{\textsf{PRIMA}^+}
\newcommand{\nodeselect}{\textit{NodeSelection}}

\newcommand{\dbBook}{\mbox{Douban-Book}\xspace}
\newcommand{\dbMovie}{\mbox{Douban-Movie}\xspace}
\newcommand{\net}{\mbox{NetHEPT}\xspace}
\newcommand{\twit}{\mbox{Twitter}\xspace}
\newcommand{\lfg}{\mbox{LastfmGenres}\xspace}
\newcommand{\orkut}{\mbox{Orkut}\xspace}
\newcommand{\db}{{Douban\mbox{-}Book}\xspace}
\newcommand{\dm}{{Douban\mbox{-}Movie}\xspace}

\newcommand{\rr}{Round-robin}
\newcommand{\sn}{Snake}

\newcommand{\ind}{d_{in}}

\begin{document}

% ****************** TITLE ****************************************
\title{Maximizing Social Welfare in a Competitive Diffusion Model} 

\numberofauthors{1}
\author{
\begin{tabular}{ccc}
Prithu Banerjee$^\dag$ \hspace{2mm} & Wei Chen$^\ddag$  \hspace{2mm} & Laks V.S. Lakshmanan$^\dag$ \hspace{2mm} \\ %& Wei Lu$^\amalg$ \\
\end{tabular}
\\$ $\\
\begin{tabular}{cc}
$^\dag$\affaddr{University of British Columbia, {\sf \{prithu,laks\}@cs.ubc.ca}}  & $^\ddag$\affaddr{Microsoft Research, {\sf weic@microsoft.com}} %\\ %&$^\amalg$\affaddr{LinkedIn}\\
%\affaddr{Vancouver, B.C., Canada}  &\affaddr{Beijing, China} & \affaddr{Sunnyvale, CA} \\
%{\sf \{prithu,laks\}@cs.ubc.ca} & {\sf weic@microsoft.com} %& {\sf welu@cs.ubc.ca}
\end{tabular}
}

%\renewcommand{\textrightarrow}{$\rightarrow$}

%\input{sec-abstract}
%%%%%%%%%%%%%%%%%%%%%%%%%%%%%%%%%%%
\eat{


}
%%%%%%%%%%%%%%%%%%%%%%%%%%%%%%%%%%%%%%%%%%%%%%%%%%%%%%%%%%
%\keywords{Influence Maximization, Non-submodular Maximization}

\maketitle

%\weic{I added ``Model'' in the title. Otherwise, it reads strange to me.}
\begin{abstract}

Influence maximization (IM) has garnered a lot of attention in the literature owing to applications such as viral marketing and infection containment.
It aims to select a small number of seed users to adopt an item such that adoption propagates to a large number of users in the network. Competitive IM focuses on the propagation of competing items in the network. Existing works on competitive IM have several limitations. (1) They fail to incorporate economic incentives in users' decision making in item adoptions. (2) Majority of the works aim to maximize the adoption of one particular item, and ignore the collective role that different items play.  (3) They focus mostly on one aspect of competition -- pure competition. To address these concerns we study competitive IM under a utility-driven propagation model called UIC, and study social welfare maximization. The problem in general is not only NP-hard but also NP-hard to approximate within any constant factor. We, therefore, devise instant dependent efficient approximation algorithms for the general case as well as a $(1-1/e-\epsilon)$-approximation algorithm for a restricted setting. Our algorithms outperform different baselines on competitive IM, both in terms of solution quality and running time on large real networks under both synthetic and real utility configurations.
\end{abstract}
\vspace{-3mm} 
\section{Introduction}\label{sec:intro}
Influence maximization (IM) on social and information networks is a well-studied problem that has gained a lot of traction since it was introduced by Kempe et al. \cite{kempe03}. 
Given a network, modeled as a probabilistic graph where users are represented by nodes and their connections by edges, the problem is to identify a small set of $k$ seed nodes, such that by starting a campaign from those nodes, the expected number of users who will be influenced by the campaign, termed influence spread, is maximized. Here, the expectation is w.r.t. an underlying stochastic diffusion model that governs how the influence propagates from one node to another. 
The ``item'' being promoted by the campaign may be a product, a digital good, an innovative idea, or an opinion. 

%\note[Laks]{Two comments. 1. IC/LT by themselves do not limit us to single or multiple items. They only prescribe how influence/information propagates through the network. 2. I'd begin by saying existing works on IM have addressed both single-item diffusion and multiple-item diffusion. Then say that previous work largely focuses on pure competition between items propagating through a network. The pure competition papers typically focus on two items, assume that seeds of one of them are given and address seed selecting for the other item (the follower item). What is missing is soft competition and competition among $m>2$ items. While saying what is missing, say why it is important, i.e., motivate it. \red{addressed}} 
%Some exceptions include the few complementarity papers. What is missing is soft competition. 
Existing works on IM typically focus on two types of diffusion models --  \textit{single item} diffusion and diffusion of \textit{multiple items under pure competition}. The two classic diffusion models, Independent Cascade (IC) and Linear Threshold (LT), were proposed in \cite{kempe03}. Advances on these lines of research have led to better scalable approximation algorithms and heuristics \cite{ChenWW10,xiaokui-opim-sigmod-2018,tang15}. 
Most studies on multiple-item diffusion focus on two items in pure competition~\cite{zhu2016minimum,lu2013,infbook,PathakBS10}, that is, every node would only adopt at most one item, never both.
The typical objective is to select seeds for the second item (the follower item) to maximize its number of adoptions, or minimize the  spread of the first item \cite{infbook}. 
%\note[Laks]{Pure competition implies at most 1 item will be adopted. :-) 
%You can also say that an underlying assumption in existing works is that mere awarensss of a product by a user implies its adoption by the user. Say how you differ and why. \red{addressed}}  

\eat{ 
\weic{The previous version starts with talking about the limitation of pure competition, which I think is minor comparing to utility-driven approach and social welfare consideration. The example of iphone etc. is alo long, less focused. 
Overall, I feel that the previous version still does not address Laks' concern well, so I am restructuring the intro from this point on.}
} 

There are a number of key issues on multiple item diffusion that are not satisfactorily addressed in most prior studies.
First, most propagation models are purely stochastic, in which if a node $v$ is influenced by a neighboring node $u$ on certain item, it will either deterministically or probabilistically  adopt the item, without any consideration of the utility of that item for the node. This fails to incorporate  
	economic incentives into the user adoption behavior.
Second, most studies focus on \textit{pure}  competition, where each node adopts at most one item, and ignore the possibility of  nodes adopting multiple items. \eat{during the propagation.} For instance, when items are involved in a partial competition, their combined utility may still be more than the individual utility, although it may be less than the sum of their utilities. 
Third, most studies on competition focus on the objective of maximizing the influence of one item given other items, or minimizing the influence of existing items, and {\color{black}do not consider maximizing the overall welfare caused by all item adoptions.}

The study by Banerjee et al. \cite{ban2019} is  unique in addressing the above issues.
It proposes the utility-based independent cascade model UIC, in which: (a) each item has a utility determined by its value, price and a noise term, and each node selects the best item or itemset that offers the highest utility among all items that the node becomes aware of thanks to its neighbors' influence;  and (b) the utility-based adoption naturally \eat{incorporates} models the adoption of multiple items, \eat{either as complementary items or as items of partial competition} in a framework that allows arbitrary interactions between items, based on chosen value functions. Banerjee et al. \cite{ban2019} 
\eat{The authors} study the maximization of expected social welfare, defined as the the total sum of the utilities of items adopted by all network nodes, in expectation.
However, their study \eat{paper only  study social welfare maximization in} is confined to  the {\sl complementary} item scenario, where item utilities increase when bundled together.

In this paper, we complement the study in \cite{ban2019} by considering the \emph{social welfare maximization} problem in the UIC model when items are purely or partially \emph{competitive.} Partial (pure) competition means adopting an item makes a user less likely (resp., impossible) to adopt another item. To motivate the problem, we note that for a social network platform owner (also called the \textit{host}), one natural objective might be to optimize the advertising revenue, as studied by   Chalermsook et al. \cite{chalermsook2015social}, or a proxy thereof, such as expected number of item adoptions. On the other hand, one of the key  assets of a network host is the loyalty and engagement of its user base, on which the host relies for its revenue from advertising and other means. Thus, while launching campaigns, it is \eat{in the best interest of} equally natural for  the host to take into account users' satisfaction by making users aware of itemsets that increase their utility. Social Welfare, being the sum of utilities of itemsets adopted by users, is directly in line with this objective. 

\eat{Below, we argue that maximizing the social welfare under the competitive setting is a natural objective for the host \cite{chalermsook2015social}.} As a real application, consider a music streaming platform such as the Last.fm. Benson et al. \cite{benson2018discrete} using their discrete choice model showed existence of competition across different genres of songs in the Last.fm dataset. In a platform such as Last.fm, the platform owner (i.e., host) completely controls the promotion of songs and the host would like to  keep making engaging recommendations to the users. Even when there are multiple competing songs from different genres, the host should recommend based on users' preferences, i.e., the users' utility. A similar idea extends to different competing products that an e-retailer like Amazon sells directly. Those products are already procured by the e-retailer  and it has full control over how it wants to sell them. \eat{Amazon would not care much about the individual sale count of these products. Instead,} 
Once again, in this setting, keeping users' satisfaction from adopting these products high helps maintain a loyal and engaged user base. 
\eat{Such real-world scenarios clearly demonstrate that there exists platforms that would provide seeding service for potentially competing companies or campaigns, but the objective of the platform is to maximize the overall satisfaction of all users on the platform, so that they stay engaged and remain loyal.} Thus maximizing the overall social welfare is in line with the goal of the platform. While \cite{ban2019} studies this problem for complementary items, social welfare maximization under competing products is open. Moreover, under pure competition, the bundling algorithm of \cite{ban2019} would lead to nodes adopting at most one of several  competing items, leading to poor social welfare.

Compared to \cite{ban2019}, we also consider a more flexible setting where the allocation of some items has been fixed (e.g., the items had the seeds selected by the host earlier) and the host is only 
	allocating seeds for the remaining items. Once again, the objective is still to maximize the total social welfare of all users in the network.
We call this the \prob problem (for Competitive Welfare Maximization).

%\note[Laks]{The paragraph above needs work. It is not compatible with our overall story that the host always chooses the seeds.}
As it turns out, \prob under UIC is \emph{significantly more difficult than the welfare maximization problem in the complementary setting studied in \cite{ban2019}.} 
We show that when treating the allocation as a set of item-node pairs, the welfare objective function is neither monotone nor submodular.
Moreover, with a non-trivial reduction, we prove that \prob is in general NP-hard to  approximate to within any constant factor. {\color{black}In contrast a constant approximation was possible in the setting considered in \cite{ban2019}.}

Despite all these difficulties, we design several algorithms that either provide an instance-dependent approximation guarantee in the general case, or
	better (constant) approximation guarantee in some special cases.
In particular, we first design  algorithm \sgrd  which provides a $\frac{\umin}{\umax}(1-\frac{1}{e}-\epsilon)$-approximation guarantee for the general \prob setting, where
	$\umin$ is the minimum expected utility among all individual items, $\umax$ is the expected maximum utility among all item bundles, and
	$\epsilon>0$ is any small positive number. 
Next, when the fixed itemset is empty, we complement \sgrd with \mgrd, which guarantees $\frac{1}{m}(1-\frac{1}{e}-\epsilon)$-approximation, where $m$ is the total number of items.
Thus, when \sgrd and \mgrd work together, we can guarantee $\max(\frac{\umin}{\umax},\frac{1}{m})(1-\frac{1}{e}-\epsilon)$-approximation when there are no prior allocated items.
We can see that when the utility difference among items is not high or the number of items is small, the above algorithms can achieve a reasonable approximation performance.
Finally, in the special case where we have a unique superior item with utility better than all other items, all other items have had their allocations fixed, and items exhibit pure competition, we design
	an efficient algorithm that achieves $(1-\frac{1}{e}-\epsilon)$-approximation.

We extensively test our algorithms against state-of-the-art IM algorithms {\color{black}under seven different utility configurations including both real and synthetic ones, which capture different aspects of competition.} Our results on real networks show that our algorithms produce social welfare up to five times higher than the baselines. Furthermore, they easily scale to large networks with millions of nodes and billions of edges. {\color{black} We also empirically test the effect of social welfare maximization on adoption count and show that whereas the overall adoption count remains the same, social welfare is maximized by reducing adoption  of just the inferior items.} To summarize, our major contributions are as follows:
\vspace{-2.0 mm}
\begin{itemize}[noitemsep]  
\item We are the first to study the competitive social welfare maximization problem \prob under the utility-based UIC model (\textsection \ref{sec:model}).

\item We show that  social welfare is neither monotone, submodular, nor supermodular; furthermore, it is NP-hard to approximate  \prob within any constant factor, in general (\textsection \ref{sec:prop}). \eat{by designing non-trivial gadget that highlights the specific challenges in approximating the problem} 
%The proof is highly non-trivial and requires construction of complex network structure and utility configuration. 

\item 
We provide several algorithms that either solve the \prob in the general setting with a utility-dependent approximation guarantee, or have better (constant) approximation guarantees in special cases (\textsection\ref{sec:algo}).

\item We conducted an extensive experimental evaluation over several real social networks comparing our algorithms with existing algorithms. Our results show that our algorithms significantly dominate existing algorithms and validate that our algorithms both deliver good quality and  scale to large networks (\textsection\ref{sec:exp}).
%
%In \textsection \ref{sec:algo}, for the general \prob problem we first give a utility dependent approximation algorithm. Then under some restricted version of the problem we design a constant $(1-\frac{1}{e})$ approximation algorithm. Along the way we first present a prefix preserving on the marginal property and extend state of the art IMM algorithm to have that property. We call this algorithm $\PRIMAP$. We also extend the RR-set machinery by adding a notion of weight, to directly work for social welfare objective. 
%\note[Laks]{Not sure if showing this is worth being highlighted as a contribution. You should judge based on novelty and tech. depth. \red{addressed}} 
%
%\item Our algorithms are rigorously tested on five real world networks in \textsection \ref{sec:exp} against state-of-the-art IM algorithm for competition. Under various utility configurations, our algorithms constantly outperforms the baseline both on social welfare and running time.  
\end{itemize} 
\vspace{-2mm}
Background and related work are discussed in \textsection\ref{sec:related}. We conclude the paper and discuss future work in \textsection\ref{sec:concl}. 
Proofs compressed or omitted for lack of space can be found in \cite{cepic-arxiv}. 

%\note[Laks]{See if we need to revise the list of contributions for the revision.}
%\note[Laks]{There may be a need for a couple of more passes after you incorporate these comments. IOW, right now, the intro. has (most of) the needed raw material. Presentation needs work.\red{hopefully getting addressed :)}}

\vspace{-1mm} 
\section{Background \& Related Work}\label{sec:related}

\noindent 
{\bf One Item IM}: 
A directed graph $G=(V,E,p)$ represents a social network with users $V$ and a set of connections (edges) $E$. The function $p: E \to [0,1]$ specifies influence probabilities between users. 
Independent cascade (IC) model is a commonly used discrete time diffusion model~\cite{kempe03,infbook}. 
Given a seed set $S\subset V$, at time $t=0$, only the seed nodes in $S$ are active. For $t>0$, if a node $u$ becomes active at $t-1$, then it makes one attempt to activate its every inactive out-neighbor $v$, with
	success probability $p_{uv} := p(u,v)$. The diffusion stops when no more nodes can become active. 

For a seed set $S\subset V$, we use $\sigma(S)$ to denote the \emph{influence spread} of $S$, i.e., the expected number of active nodes at the end of  diffusion from $S$. 
For a seed budget $k$ and a diffusion model, {\em influence maximization} (IM) problem is to find a seed set $S\subset V$ with $|S| \leq k$
	such that the influence spread $\sigma(S)$ under the model is maximized~\cite{kempe03}. 
 
\eat{A few important properties of a set function are often considered while designing IM solutions.} A set function $f: 2^V \to \mathbb{R}$
is \emph{monotone} if $f(S) \leq f(T)$ whenever $S\subseteq T\subseteq V$; \emph{submodular} if for any $S\subseteq T\subseteq V$ and any $x \in V \setminus T$, $f(S\cup\{x\}) - f(S) \geq f(T\cup\{x\}) - f(T)$; $f$ is \emph{supermodular} if $-f$ is submodular; and $f$ is \emph{modular} if it is both submodular and supermodular.

Under the IC model, IM is intractable \cite{kempe03, ChenWW10, ChenWW10b}. However 
$\sigma(\cdot)$ is {\em monotone} and {\em submodular}. Hence, using Monte Carlo simulation for estimating the spread, a simple greedy algorithm delivers a $(1-1/e-\epsilon)$-approximation to the optimal solution, for any $\epsilon > 0$~\cite{kempe03, kapraov-etal-greedy-opt-soda-2013,jung2012}. 
The concept of reverse reachable (RR) sets proposed by Borgs et al. \cite{borgs14}, has led to 
a family of scalable state-of-the-art approximation algorithms such as IMM and SSA for IM~\cite{tang15,Nguyen2016,Huang2017,chen-etal-pvldb-2015,li-etal-pvldb-2015}.

\noindent 
{\bf Multiple item competitive IM}: More recently, IM has been studied involving 
independent %(thus, non-competing) 
items~\cite{dattaMS10}, %later work studied 
and competing items  
~\cite{HeSCJ12,BudakAA11,lu2013,zhu2016minimum,BharathiKS07}. In \cite{lin2015analyzing} authors studied the problem under pure competition, whereas \cite{garimella2017balancing} aims to maximize balanced exposure in the network in presence of two competing ideas,{\color{black} and \cite{lu2013bang} ensured fairness in the adoption of competing items. These works, however, are restricted to specific type of competition.}
The \comic model proposed by Lu et al.~\cite{lu2015arxiv} can model any arbitrary degree of interaction between a pair of items. Their main study is therefore restricted to the diffusion of two items. {\color{black} \cite{li2019maximizing} looks into different facets of items to compute influence (e.g., topics of documents).
However, unlike our work, they do not consider item utility in adoption decisions made by users. Furthermore, their objective function is based on traditional (expected) number of item adoptions.
In addition to the above  differences, our objective is to maximize the social welfare that none of these papers have studied.} 
However, unlike our work, they do not consider item utility in adoption decisions made by users. Furthermore, their objective function is based on traditional (expected) number of item adoptions. 
In addition to the above  differences, our objective is to maximize the social welfare that none of these papers have studied.
For a more comprehensive survey on competitive influence models, see \cite{infbook,li-etal-im-survey-tkde-2018}.

\noindent 
{\bf Social welfare maximization}: 
Utility driven adoptions have been studied in economics~\cite{myerson1981optimal, nisan2007,AbramowitzA18,boadway1984welfare}. Given items and users, and the utility functions of users for various subsets of items, the problem is to find an allocation of items to users such that the sum of utilities of users, is maximized. Since the problem is intractable, approximation algorithms have been developed \cite{feige-vondrak-demand-2010, kapraov-etal-greedy-opt-soda-2013, korula-etal-online-swm-arxiv-2017}. {\color{black} \cite{benson2018discrete} proposed a discrete choice model to learn the utilities of itemsets from the users' adoption logs. Learning utility is complementary to our problem. Moreover none of these works consider a social network and the effect of recursive propagation of item adoptions by its users.}

{\color{black} Host's perspective in the context of IM have been studied. \cite{aslay2017revenue} directly maximizes the revenue earned by a network host, whereas \cite{aslay2015viral} aimed to minimize the regret of seed selection. These works donot consider the overall social welfare. Utility based adoption decisions of users are also not part of their formalism.}
Welfare maximization on social networks has  been studied in a few recent papers\cite{BhattacharyaDHS17,SunCLWSZL11}.
%%%%%%%%%%%
%removing as the connection seems weak
\eat{
Sun et al. \cite{SunCLWSZL11} study participation maximization where an item is a discussion topic, and adopting an item means posting or replying on the topic. They, however, assume that
items are independent (i.e., value function is additive rather than being supermodular or submodular), and also the budget constraints the number of items each seed node can be allocated with, rather than the number of seeds each item can be allocated to as studied in our model.
}
%%%%%%%%%%%%%
Bhattacharya et al. \cite{BhattacharyaDHS17} consider item allocations to nodes for welfare maximization in a network with network externalities. Their model does not consider the effect of \eat{viral marketing.} recursive propagation nor competition. \eat{A user's valuation of an item is affected by the network externality, i.e., the number of her direct one- or two-hop neighbors in the network adopting the same item.} In addition, they do not consider budget constraints. 
\eat{, so	 an item could be allocated to any number of nodes.}  
In contrast,  our focus is on competition,  with budget constraints on every item. 

Banerjee et al. \cite{ban2019} studied welfare maximization under viral marketing using the UIC propagation model that we also use. However, their work focus strictly on  complementary items, with supermodular value functions. As a result, their objective is monotone, and further satisfies a nice ``reachability'' property (details in \textsection\ref{sec:prop}), which paved the way for efficient approximation. {\color{black}However such complementary only setting fails to model many real world platforms where competing items are also present as highlighted in the introduction.} 
We instead focus on \emph{competing} items. Consequently, the objective  becomes not only non-monotone, non-submdular, and non-supermodular, but unlike in \cite{ban2019},  is inapproximable within any constant. In spite of this, we develop utility dependent approximation algorithms as well as a constant approximation algorithm for special cases. 

In summary, to our knowledge, {\sl our study is the first to address social welfare maximization in a network with influence propagation, competing items, and budget constraints, where item adoption is driven by utility.}

\vspace*{-1ex} 
\section{\model Model under competition}\label{sec:model}

In this section, we first briefly review the \emph{utility driven independent cascade} model (UIC for short) proposed in \cite{ban2019}. %Unlike \cite{ban2019}, which studied the complementary setting of UIC, our focus is on competitive setting. 
Then we describe the competitive setting of UIC studied in this paper and formally state the new problem we address. 
 
\noindent
{\bf Review of UIC  Model:} 
UIC integrates utility driven adoption decision of nodes, with item propagation. Every node has two sets of items -- \textit{desire set} and \textit{adoption set}. Desire set is the set of items that the node has been informed about (and thus potentially desires), via propagation or seeding. Adoption set, is the subset of the desire set that has the highest utility, and is adopted by the user. The utility of an itemset $\itemset \subseteq \allitems$ is derived as $\util(\itemset) = \val(\itemset) - \price(\itemset) + \noise(\itemset)$, where $\val(\cdot)$ denotes users' latent valuation for an itemset, $\price(\cdot)$ denotes the price that user needs to pay, and $\noise(\cdot)$ is a random noise term that denotes our uncertainty in users' valuation.  

%Adoption is progressive, i.e., once a node adopts an item, it cannot unadopt it later. 

Budget vector $\vec{b} = (b_1, ..., b_{|\allitems|})$ represents the budgets associated with the items, i.e., the number of seed nodes that can be allocated with that item. An \emph{allocation} is a relation $\allalloc \subset V \times \allitems$ such that $\forall i\in\allitems: |\{(v,i)\mid v\in V\}| \le b_i$. $S_i^{\allalloc} := \{v \mid (v,i) \in \allalloc\}$ denotes the \emph{seed nodes} of $\allalloc$ for item $i$ and $S^{\allalloc} := \bigcup_{i\in\allitems} S_i^{\allalloc}$. When the  allocation $\allalloc$ is clear from the context, we write $S$ (resp., $S_i$) to denote $S^{\allalloc}$ (resp., $S_i^{\allalloc}$). 

Before a diffusion begins, the noise terms of all items are sampled,  and they are used until the end of that diffusion. The diffusion proceeds in discrete time steps, starting from $t=1$. $\awares(v,t)$ and $\adopts(v,t)$ denote the desire and adoption sets of node $v$ at time $t$. 
At $t=1$, the seed nodes have their desire sets initialized according to the allocation $\allalloc$ as,  $\awares(v,1) = \{i \mid (v,i) \in \allalloc\}$, $\forall v \in S^{\allalloc}$. These seed nodes then adopt the subset of items from the desire set that 
	maximizes the utility.
The propagation then unfolds recursively for $t \geq 2$ in the following way.
\eat{Nodes adopt . }
Once a node $u'$ adopts an item $i$ at time $t-1$, it influences its out-neighbor $u$ 
	with probability $p_{u'u}$, and if
	it succeeds, then $i$ is added to the desire set of $u$ at time $t$. Subsequently $u$ adopts the subset of items from the desire set of $u$ that 
	maximizes the utility. Adoption is progressive, i.e., once a node adopts an item, it cannot unadopt it later. Thus $\adopts(\user,t) = \argmax_{T \subseteq \awares(\user,t)} \{\util(T) \mid T \supseteq \adopts(\user,t-1)\;\wedge\;\util(T)\ge 0\}$. The propagation converges when there is no new adoption in the network.
	\eat{$\adopts(\user)$ denotes the adoption set of user $\user$ upon convergence.}  
	For more details readers are referred to \cite{ban2019}.

\noindent
{\bf Social welfare maximization relative to a fixed seed set:\ \ } 
Let $G = (V,E,p)$ be a social network, $\allitems$ the universe of  items under consideration.    We consider a utility-based objective called {\em social welfare}, which is the sum of all users' utilities of itemsets adopted by them after the propagation converges. Formally, $\E[\util(\adopt^{\allalloc}(u))]$ is the expected utility that a user $\user$ attains for a seed allocation $\allalloc$ after the propagation ends. The \emph{expected social welfare} %(termed ``consumer surplus'' in algorithmic game theory) 
for $\allalloc$,  is $\rho(\allalloc) = \sum_{u \in V}\E[\util(\adopt^{\allalloc}(\user))]$, where the expectation is over both the randomness of propagation and randomness of noise terms $\noise(.)$

In a social network, a campaign may often be launched on top of other existing campaigns, where the seeds for some items  $I_1 \subset \allitems$ may already be fixed. Let  $\allallocp$ be this fixed allocation for items in $I_1$.  Then $I_2 = \allitems \setminus I_1$ is the set of items for which the seeds are to be selected.
We define the problem of maximizing expected social welfare, on top of a fixed seed allocation as follows. 
%\vspace{-3mm}

\noindent
{\bf Welfare maximization under competition:\ \ } 
\eat{Social welfare measures network users' collective satisfaction. Thus, even when items are competing, it is natural for a host to maximize welfare so that the users of the network are happy and  stay engaged in the network.} To model competition, we assume that $\val$ is submodular \cite{Carbaugh16},
  i.e., the marginal value of an item with respect to an itemset $I \subset \allitems$ decreases as $I$ grows. We assume 
$\val$ is monotone, since it is a natural property for valuations.  
We set $\val(\emptyset) = 0$. For $i \in \allitems$, $\noise(i) \sim \mathcal{D}_i$ denotes the noise term associated with item $i$, 
where the noise may be drawn from any distribution $\mathcal{D}_i$ having a zero mean. Every item has an independent noise distribution. For a set of items $I \subseteq \allitems$, we assume the noise and price to be additive. 
Since noise is drawn from a zero mean distribution, $\mathbb{E}[\util(I)] = \val(\itemset) - \price(\itemset)$. Below, we refer to $\val, \price, \{\mathcal{D}_i\}_{i\in \allitems}$, as the model parameters and denote them collectively as $\parameterset$. 
\vspace{-1mm}
\begin{problem}
[\prob]
Given $G = (V,E,p)$, the set of model parameters $\parameterset$, an existing fixed allocation $\allallocp$, and budget vector $\bvec$, find a seed allocation $\allalloc^*$ for items $I_2$, such that $\forall i \in I_2$, $|\alliseeds^{\allalloc^*}| \leq b_i$ and $\allalloc^*$ maximizes the expected social welfare, i.e., $\allalloc^* = \argmax_{\allalloc} \rho(\allalloc \cup \allallocp)$.
\end{problem}
\vspace{-1mm}
Note that this problem subsumes the typical "fresh campaigns"  setting as a special case where $I_1 = \emptyset$ (and hence $\allalloc^p = \emptyset$).

\eat{ 
\note[Laks]{{\bf Remarks}: 1. The above problem formulation is applicable regardless of the nature of interaction between items, i.e., complementary, partially competitive, purely competitive, or independent. The properties of the corresponding social welfare functions may differ, depending on the nature of the interaction. 2. We said that we should include a brief motivation for studying SW in a competitive setting.}
} 

\noindent
{\bf An equivalent possible world model: } 
In \cite{ban2019}, the authors proposed an equivalent possible world interpretation of the diffusion under UIC, which we will find useful. We briefly review this  below.  
\eat{ 
Here we present the equivalence property that we will use to prove our results in this paper.} 
Let $\langle G, {\sf Param}\rangle$ be an instance of \prob, where $G=(V,E,p)$. A \emph{possible world} $w = (w_1, w_2)$, consists an {\em edge possible world} (edge world) $w_1$, and a {\em noise possible world} (noise world) $w_2$:
	$w_1$ is a deterministic graph sampled from the distribution associated with $G$, where each edge $(u,v)\in E$ is sampled  in with an independent
	probability of $p_{uv}$; and $w_2$ is a sample of noise terms for items in $\allitems$, drawn from noise distributions in {\sf Param}. 
Note that propagation and adoption in $w$ is fully deterministic. In a possible world $w$, $\noise_w(i)$ is the noise for item $i$ and $\util_w(I)$ is the (deterministic) utility of itemset $I$. 
\eat{Given an allocation $\allalloc$, the desire and adoption sets of a node $u$ at time $t$ in world $w$ are $\awarews(u,t)$ and $\adoptws(u,t)$ respectively. } 
The \emph{social welfare} of an allocation $\allalloc$ in $w$ is $\rho_w(\allalloc) := \sum_{v\in V} \util(\adoptws(v))$, where $\adoptws(v)$ is the adoption set of $v$ at the end of the propagation in world $w$. 
The \emph{expected social welfare} of an allocation $\allalloc$ is $\rho(\allalloc) := \mathbb{E}_w[\rho_w(\allalloc)] = \mathbb{E}_{w_1}[\mathbb{E}_{w_2}[\rho_w(\allalloc)]] = \mathbb{E}_{w_2}[\mathbb{E}_{w_1}[\rho_w(\allalloc)]]$.

\vspace{-1mm} 
%\vspace{-5mm}
\section{Properties of \model}\label{sec:prop}
It is easy to see that \prob is NP-hard. 

\begin{proposition}
\prob in the \model model is NP-hard. 
\end{proposition}

\eat{
\begin{proof}
We show that Influence Maximization under the IC model, a well-known NP hard problem \cite{kempe03}, is a special case of \prob: let $\allitems = \{i\}$, set $\val(i) = 1$, $\price(i) = 0$ and set the noise term for item $i$ to $0$. This makes $\util(i) = 1$, so any influenced node will be in desire of $i$ and then adopt $i$. Thus, an allocation maximizes the expected social welfare iff the corresponding seed set  maximizes the expected spread. 
\end{proof}
 } 
 \begin{proof}[Sketch]  
 Classic IM is a special case of \prob. 
 \end{proof} 
 
Given the hardness, we examine whether social welfare satisfies monotonicity, submodularity or supermodularity. 

\eat{we explore properties of the welfare function --  monotonicity and submodularity, which can help us design efficient approximation algorithms. We begin with an equivalent possible world model semantics of \model to help our analysis. }

%%%%%%
%needs paraphrasing{

%\note[Laks]{Is is true that the equiv. PW here is the same as in [2]?}

\spara{Item blocking}
Under the \emph{complementary setting} 
in \cite{ban2019} leveraged the reachability property:  if a node $v$ adopts an item $i$ in any possible world $w$, then all the other nodes that are reachable from $v$ in $w$ will also adopt $i$. This property does \emph{not} hold under the competitive setting. In fact,  adoption of one particular item can block the propagation of another item, making social welfare non-monotone and non-submodular.  

\begin{restatable}{theorem}{thmnotsubsuper} \label{thm:prop}
Expected social welfare is not monotone, and neither submodular nor supermodular, with respect to  sets of node-item allocation pairs.
\end{restatable}

\begin{proof}
We show a counterexample for each of the three properties. Consider a simple network with two nodes $u$ and $v$, and a  directed edge $(u, v)$ with probability $1$. Assume that there is no noise, i.e., noise is $0$. There are three items in propagation whose utility configuration is shown in Fig. \ref{fig:util_tables} (a). 

\begin{figure*}[ht]
    \centering
    \includegraphics[width=1\textwidth]{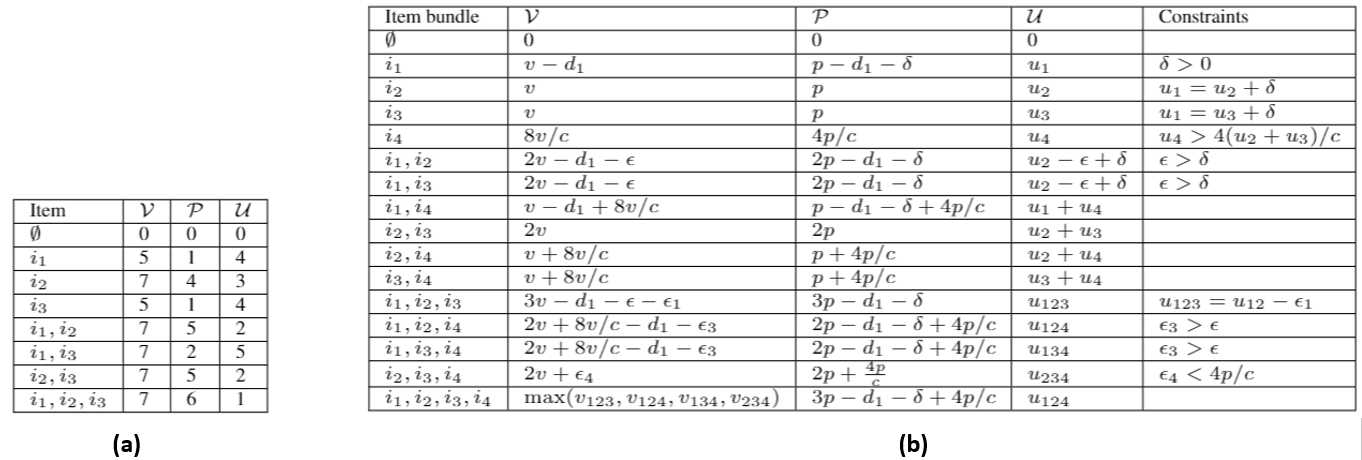}
    \caption{Utility configurations: (a) Used in Theorem 1;  (b) Used in Theorem 2}
    \label{fig:util_tables}
\end{figure*}

\spara{Monotonicity}
Consider two allocations $\allalloc^1 = \{(u,\1)\}$ and $\allalloc^2 = \{(u,\1), (v,\2)\}$. Clearly $\allalloc^1 \subset \allalloc^2$. Under $\allalloc^1$, both $u$ and $v$ adopt $\1$, thus $\rho(\allalloc^1) = 8$. However under $\allalloc^2$, $u$ adopts $\1$ but $v$ adopts $\2$. Thus $\rho(\allalloc^2) = 7 < \rho(\allalloc^1)$.

\spara{Submodularity}
Consider $\allalloc^1 = \{(v, \2)\}$, $\allalloc^2 = \{(v, \2), (v, \3)\}$ and $(u, \1)$. Clearly $\allalloc^1 \subset \allalloc^2$ and $(u, \1) \notin \allalloc^2$. Under $\allalloc^1$, only $v$ adopts $\2$. Under $\allalloc^1 \cup \{(u, \1)\}$, $u$ adopts $\1$ and $v$ adopts $\2$. So $\rho(\allalloc^1 \cup \{(u, \1)\}) - \rho(\allalloc^1) = 4$. Under $\allalloc^2$, $v$ adopts $\3$. Under $\allalloc^2 \cup \{(u, \1)\}$, $u$ adopts $\1$ and $v$ adopts $\1$ and $\3$. So $\rho(\allalloc^2 \cup \{(u, \1)\}) - \rho(\allalloc^2) = 5 > \rho(\allalloc^1 \cup \{(u, \1)\}) - \rho(\allalloc^1)$. 

\spara{Supermodularity}
Consider $\allalloc^1 = \emptyset$, $\allalloc^2 = \{(v, \2)\}$ and $(u, \1)$. Clearly $\allalloc^1 \subset \allalloc^2$ and $(u, \1) \notin \allalloc^2$. Under $\allalloc^1$, there is no adoption by any node. Under $\allalloc^1 \cup \{(u, \1)\}$, $u$ and $v$ both adopt $\1$. So $\rho(\allalloc^1 \cup \{(u, \1)\}) - \rho(\allalloc^1) = 8$. Under $\allalloc^2$, $v$ adopts $\2$. Under $\allalloc^2 \cup \{(u, \1)\}$, $u$ adopts $\1$ and $v$ adopts $\2$ . So $\rho(\allalloc^2 \cup \{(u, \1)\}) - \rho(\allalloc^2) = 4 < \rho(\allalloc^1 \cup \{(u, \1)\}) - \rho(\allalloc^1)$. 
\end{proof}

The absence  of these properties makes \prob really hard to approximate, as shown next. \eat{In fact we  prove next that there is no $PTIME$ approximation algorithm for the general version of \prob.}

\begin{theorem}
\prob in the \model model is NP-hard. Further there is no $PTIME$ algorithm that can approximate \prob within any constant factor $c$, $0<c \leq 1$, unless P = NP.
\end{theorem}

\begin{proof}

\textbf{NP-hardness}

We show that 
Influence maximization under the IC model, an
	NP hard problem, is a special case
	of \prob.

The result follows from the fact that the IM problem under the IC model is a special case of \prob: let $\allitems = \{i\}$, set $\val(i) = 1$, $\price(i) = 0$ and set the noise term for item $i$ to $0$. This makes $\util(i) = 1$ so any influenced node will adopt $i$. Thus, the expected social welfare is simply the expected spread. We know maximizing expected spread under the IC model is NP-hard \cite{kempe03}.

\textbf{Hardness of approximation}

%%%%%%%%%%%%%%%%%%% 
% TO SAVE SPACE, COMMENTING OUT MST OF THE HARDNESS OF APPROX. PROOF. NEED TO RESURRECT IT FOR THE FULL VERSION. 
%%%%%%%%%%%%%%%%%%% 

%[Sketch]
We prove the theorem by a gap introducing reduction from SET COVER. Suppose there is a PTIME $c$-approximation algorithm $\calA$ for \prob, for some $0 < c \leq 1$. \eat{Given any instance $\inst$ of SET COVER, in polynomial time we can produce a transformed instance $\calJ$ of \prob.} \eat{such that by running $\calA$ on $\calJ$, we can decide if $\inst$ is a YES-instance or a NO-instance, which is not possible unless $P=NP$.} Given an instance $\inst = (\calf, \ground)$ of SET COVER, where $\calf  = \{S_1, ..., S_r\}$ is a collection of subsets over a set of ground elements $\ground = \{g_1, ..., g_n\}$, and a number $k$ ($k<r<n$), the question is whether there exist $k$ subsets from $\calf$ that cover all the ground elements, i.e., whether $\exists \cover \subset \calf: |\cover| = k$ and $\cup_{S\in\cover}^k S = \ground$. We can transform $\inst$ in polynomial time to an instance $\calJ$ of \prob.

As an overview, our reduction will show that for a YES-instance of  SET COVER, the optimal expected welfare in the corresponding \prob instance is high and for a NO-instance, it is low. More precisely, let $x^*_y$ (resp., $x_n^*$) be the optimal welfare on the transformed instance $\calJ$ whenever the given instance $\inst$ is a YES-instance (resp., NO-instance). Our reduction ensures that $x^*_n < cx^*_y$. In this case, running $\calA$ on $\calJ$ will clearly allow us to decide if $\inst$ is a YES-instance or not, which is impossible unless P = NP. 

For the rest of the discussion we assume no noise, i.e., noise distribution has $0$ mean and variance. Also all the edge probabilities of the graph are set to $1$. The details of the reduction follow. 

\noindent{\bf Value, price and utility}: 
We consider four items \eat{in propagation} -- $\1,\2,\3$ and $\4$,  with the following utility configuration: $\1$ competes with $\2$ and $\3$, and $\1$ has a higher individual utility than both. However $\{\2,\3\}$ as a bundle has higher utility than $\1$. 
Item $\4$ has a very high utility, much higher than that of any other individual item. A node adopting $\1$ adopts $\4$ if it arrives later. However if a node adopts the bundle $\{\2,\3\}$, then it will not adopt $\4$ later. We use this configuration in the following way. For a YES-instance, $\1$ blocks $\2$ and $\3$, consequently allowing a large number of nodes to adopt $\4$. For a NO-instance, however, most nodes adopt $\{\2,\3\}$, blocking $\4$ adoption. 
Thus by setting $ c \cdot \util(\4) >  \util(\{\2,\3\})$, the desired gap in the optimal welfare is achieved. Lastly, as we will see later in the proof that we need, $ \util(\{i_2, i_3\}) < c/4 \cdot \util(\{i_1, i_4\})$.

Assuming no noise terms, Fig. \ref{fig:util_tables} (b) provides an abstract summary of this utility configuration, focusing on the items $\{\1,\2,\3, \4\}$. Note that in addition to the aforementioned constraints, the value function is  monotone and submodular, as required. 
%%%% TO RESURRECT FOR FULL VERSION 
Further, we give one such complete configuration (over all four items $\1,\2,\3,\4$) in Table \ref{tab:utility-table}, for $c=0.4$.

\noindent{\bf The network}: 
The graph instance constructed from the given instance of SET COVER is illustrated in Fig. \ref{fig:inapprox}(a). 
We first create a bipartite graph having two partitions of $r$ nodes $\{s_1, ..., s_r\}$  corresponding to the sets $S_i$ and $n$ nodes $\{g_1, ..., g_n\}$ corresponding to the ground elements $g_j$  respectively. There is a directed edge from $s_i$ node to $g_j$ node iff  $g_j \in S_i$ in the SET COVER instance. There are also $n$ number of ``$a$'', ``$b$'', ``$e$'', ``$f$'' nodes. Node $a_i$ is connected with a directed edge to the corresponding $g_i$ node. For each $g_i$, there is an incoming (directed) edge from $a_i$ and an outgoing edge to $f_i$. Each node $b_i$ is connected to $f_i$ with a path of length $2$, i.e., $b_i\hspace{-0.75ex}\rightarrow \hspace{-0.75ex}e_i\hspace{-0.75ex}\rightarrow\hspace{-0.75ex}f_i$, where $e_i$ is the intermediate node between $b_i$ and $f_i$. This construction creates the following behavior. If all the $g$ nodes adopt $\1$ then all the $f$ nodes adopt $\1$. However if any one of the ``$g$'' nodes adopts $\2$ and all the ``$e$'' nodes adopt $\3$, then all the ``$f$'' nodes adopt $\{\2,\3\}$. The significance of this behavior will be clear in the remaining part of the proof.

For a large $N >> n$ that is a multiple of $n$, we create nodes $d_1, ..., d_N$. For $1 \leq i \leq n$, we add the edges $(f_i, d_{(i*N/n - N/n)+1})$,  ..., $(f_i, d_{i*N/n})$. This gadget helps create the gap in the welfare that we are aiming for.

We create $n$ copies of ``$j$'' nodes. Each $j_i$ is connected to $o_i$ by a directed path from $j_i$ to $o_i$ \textcolor{black}{of length $3$}, where $l_i$ and $m_i$ are the intermediate nodes. Similar to $f_i$, each $o_i$ is connected to $N/n$ ``$d$'' nodes -- $d_{(i-1)N/n+1}, d_{iN/n}$. As a preview, the ``$a$'' nodes (resp., ``$b$'' nodes, ``$j$'' nodes) will serve as seeds for item $i_2$ (resp., item $i_3$ and item $i_4$). Note that the length of the paths from ``$j$'' nodes (seeds of $\4$) to ``$d$'' nodes is $4$, while the paths from the seeds of $\2$ and $\3$ to ``$d$'' nodes are of length $3$. Thus, if $\{\2,\3\}$ are not blocked by $\1$, all the ``$d$'' nodes will adopt $\{\2,\3\}$ and cannot adopt $\4$ when it arrives later. This completes one copy of the graph, shown in Fig. \ref{fig:inapprox}(a). All edge probabilities are set to $1$. We will explain the significance of node color and the surrounding box soon.
\eat{ We will explain the rest of the construction, i.e., Fig. \ref{fig:inapprox}(b) later, but before that we describe how we exploit this structure using careful seed placement, to introduce the gap in welfare between the YES and NO instances.
} 

\noindent
{\bf Budgets and seed allocation}: We set the budgets for $\2, \3, \4$ to $n$ each. The ``$a$'' nodes are seeded with $\2$, ``$b$'' nodes are seeded with $\3$ and ``$j$'' nodes are seeded with $\4$. These seeds are fixed (see Fig.~\ref{fig:inapprox}). The budget for $\1$ is set to $k$ and these seeds are to be selected so as to maximize the expected social welfare. We complete the construction of the instance $\calJ$ of \prob by making $N$ copies of the graph described above.

Notice for YES-instance of the set cover, the $N$ number of ``$d$'' nodes adopt $\{\1,\4\}$. Hence we have,
%Before we describe the details, we have the following 

\vspace{-1mm} 
\begin{claim}
\label{claim1} 
Suppose $\inst = (\calf, \ground)$ is a YES-instance and $\calJ'$ the transformed instance of \prob corresponding to Fig.~\ref{fig:inapprox}(a) and the seed allocation described above. The optimal welfare on $\calJ'$ is $x^* > N \times \util(\{\1, \4\})$. 
\end{claim}
\vspace{-0.5mm} 
%%%% TO RESURRECT FOR FULL VERSION 

\begin{proof} 

For a YES-instance, choosing the corresponding $s$ nodes of the SET COVER solution maximizes the welfare. In this case, every ``$g$'' node has at least one in-neighbor that adopted $\1$ at time $t=1$. Thus, all ``$g$'' nodes adopt $\1$ at time $t=2$. Consequently all the ``$f$'' and ``$d$'' nodes adopt $\1$ at time $t=3$ and $t=4$ respectively. Later at $t=5$ when $\4$ arrives, those ``$d$''  nodes adopt $\{\1,\4\}$. The optimal welfare in this case $x^* > N \times \util(\{\1, \4\})$.
 
%%%% TO RESURRECT FOR FULL VERSION 
\end{proof}

%%%%%%%%%%%%%%%%%%%%%%%% DETAILED EXPLANATION/MOTIVATION -- OMITTED
 
For a NO-instance, if we hypothetically fix the seeds of $\1$ nodes to $s$ nodes, then since there are no $k$ ``$s$'' nodes that can cover all the $g$ nodes, there will be at least one $g_i$ node that will not have an in-neighbor adopting $\1$. Thus that $g_i$ node, at time $t=2$, will adopt $\2$, being influenced by the corresponding $a_i$ node. At $t=3$, since $\{\2,\3\}$ as bundle has a higher utility than $\1$, all ``$f$'' nodes adopt $\{\2,\3\}$, consequently all ``$d$'' nodes will also adopt $\{\2,\3\}$ and will not be able to adopt $\4$. Thus in this case, assuming a very large value of $N$, $x^* \leq \util(\{\2,\3\}) \times N + o(1)$. However, for a NO-instance, the optimal welfare cannot be achieved by choosing ``$s$'' nodes as seeds for $\1$. Instead, the $g$ nodes should directly be seeded with $\1$. In that case, before $\2$ arrives, at $t=1$, those $k$ seeded ``$g$'' nodes adopt $\1$. At $t=2$ all ``$f$'' nodes also adopt $\1$ and at $t=3$ all ``$d$'' nodes adopt $\1$. Later these ``$d$'' nodes adopt $\4$. Thus welfare is similar to that of YES-instance (which is undesirable). 

%%%%%%%%%%%%%%%%%%%%%%%%

\noindent
{\bf Completing construction of $\calJ$}:

To circumvent the above mentioned problem, we next create $N$ copies of the structure $\calJ'$ described above as shown in Fig. \ref{fig:inapprox}(b). Except for the ``$s$'', ``$a$'', ``$b$'' and ``$j$'' nodes, all other nodes and their connections are duplicated exactly the same way in each of those $N$ copies. The nodes that are not duplicated are colored in red. The duplicated nodes are connected to non-duplicated nodes in the same way across all the $N$ copies of the structure. E.g., across different copies, the same duplicated $g_i$ nodes are connected to the (non-duplicated) $s_i$ node, depending on whether $g_i \in S_i$ in the SET COVER instance. Similarly $j_1$ is connected to copies of $l_1$, i.e., to $l_{11}, ..., l_{1N}$. In other words the network structure of Fig. \ref{fig:inapprox}(a) i.e. enclosed in the box, is replicated $N$ times, shown using $N$ boxes in \ref{fig:inapprox}(b). 
Together with the seed allocation of items $i_2, i_3, i_4$ above, this completes the construction of instance $\calJ$ of the problem.

Now there are $N^2$ number of ``d'' nodes. Following Claim \ref{claim1} for a YES-instance of the set cover, ``d'' nodes adopt $\{\1,\4\}$. Hence,
\eat{Fixed seed nodes remains as before.}
\vspace{-1mm} 
\begin{claim}\label{claim2}
Suppose $\inst = (\calf, \ground)$ is a YES-instance and $\calJ$ the transformed instance of \prob corresponding to Fig. ~\ref{fig:inapprox}(b) and the seed allocation described above. The optimal welfare on $\calJ$ is $x^* > N^2 \times \util(\{\1, \4\})$.
\end{claim}

\begin{proof}  
 
There are $N^2$  ``$d$'' nodes in all. For a YES instance, the optimal seeds for $\1$ are exactly the solution of SET COVER. It follows from Claim~\ref{claim1} that the optimal welfare in that case is 
\begin{equation}\label{eq:wy}
x_y^* > N^2 \times \util(\{\1, \4\}).
\end{equation}
 \ \\
%%%% TO RESURRECT FOR FULL VERSION 
\end{proof}

%\vspace*{-1.5ex} 
For a NO-instance, maximum number of ``$d$'' nodes adopt $i_4$. The only candidate seeds which could achieve that are ``$s$'', ``$g$'', ``$f$'', ``$e$'', and ``$o$'' nodes. We show that regardless of which $k$ seeds are chosen for item $i_1$, the welfare achieved is $x_n^* < cN^2\util(i_4)$. First, observe that choosing $k$ ``$g$'' nodes as seeds of item $i_1$ achieves a welfare no less than that of any other choice of $k$ seeds for $i_1$.
\eat{
\note[Laks]{Do we need to elaborate on this?}} 
\eat{
Now by seeding $k$ ``$g$'' nodes with $\1$, $d$ nodes of only the $k$ of the $N$ copies can be made to adopt $\4$. For rest of the $(N-k)$ copies, as shown previously, the ``$d$'' nodes in those copies adopt $\{\2,\3\}$ and block $i_4$. Thus the optimal welfare is $x^*_n \leq kN\util(\4) + (N-k)N\util(\{\2,\3\}) +o(1)$.
\note[Laks]{Details of $o(1)$?} 

%I am providing the exact welfare for both NO and YES instances.

NO instance:
} 
In the $k$ copies where $g$ nodes are seeded with $i_1$ we have the following adoptions. 1 number of ``$g$'' and $n$ number of ``$f$'' nodes adopt $i_1$; $n-1$ number of ``$g$'' adopt $i_2$; $n$ number of ``$e$'' adopt $i_3$; $n$ number of ``$l$'', ``$m$'' and ``$o$'' nodes adopt $i_4$; and $N$ number of ``$d$'' nodes adopt $\{i_1, i_4\}$.
\eat{
Welfare $$k((n+1)\util(i_1) + (n-1)\util(i_2) + n \util(i_3) + 3n \util(i_4) + N \util(\{i_1, i_4\}) )$$.
}
For the remaining $N-k$ copies we have: $n$ number of ``$g$'' nodes adopt $i_2$; $n$ number of ``$e$'' nodes adopt $i_3$; $3n$ number of ``$l$'', ``$m$'' and ``$o$'' nodes adopt $i_4$; $N$ number of ``$d$'' nodes and $n$ number of ``$f$ nodes adopt $\{i_2, i_3\}$.
\eat{ 
The resulting welfare is
$$(N-k)(n\util(i_2) + n \util(i_3) + 3n \util(i_4) + (N+n) \util(\{i_2, i_3\}))$$
}
Lastly from the seeds, $n$ number of ``$a$'' nodes adopt $i_2$; $n$ number of ``$b$'' nodes adopt $i_3$; and $n$ number of ``$j$'' nodes adopt $i_4$.
So the total welfare for a NO-instance is: 
\vspace{-2mm}
\begin{align*}
&k[(n+1)\util(i_1) + (2n-1)\util(i_2) + 2n \util(i_3) + 4n \util(i_4) \\
&+ N \util(\{i_1, i_4\}) ]+(N-k)[(2n)\util(i_2) + (2n)\util(i_3) + 4n \util(i_4) \\
&+ (N+n) \util(\{i_2, i_3\})) +n(\util(i_2)+\util(i_3)+\util(i_4)].\\
&=(kn+k)\util(i_1) + (n - k +2Nn)\util(i_2) + (2Nn +n)\util(i_3) \\
&+ (3Nn +n)\util(i_4) + Nk\util({i_1,i_4})+(N-k)(N+n)\util({i_2,i_3}) \;(*)
\end{align*}
\vspace{-2mm}
\eat{
$$k[(n+1)\util(i_1) + (2n-1)\util(i_2) + 2n \util(i_3) + 4n \util(i_4) + N \util(\{i_1, i_4\}) ] +$$
$$(N-k)[(2n)\util(i_2) + (2n)\util(i_3) + 4n \util(i_4) + (N+n) \util(\{i_2, i_3\}))+$$
$$n(\util(i_2)+\util(i_3)+\util(i_4)] \;\;\; (*)$$
}

Since $\util(\{i_2,i_3\}) > \util(i_1) > \util(i_2) = \util(i_3)$, 
\begin{align*}
(*) &< (k+2n+5Nn+N^2-kN)\util(\{i_2,i_3\}) \\
&+(3Nn+n)\util(i_4)+Nk\util(\{i_1, i_4\})\\
&= N^2\util(\{i_2,i_3\}) + (k+2n+5Nn-kN)\util(\{i_2,i_3\} \\
&+(3Nn+n)\util(i_4)+Nk\util(\{i_1, i_4\}) \\
&< N^2\util(\{i_2,i_3\}) + (8Nn-kN)\util(\{i_2,i_3\} \\
&+4Nn\util(i_4)+Nk\util(\{i_1, i_4\}) \;\;\; (**)
\vspace{-3mm}
\end{align*}
\begin{figure}[ht]
    \centering
    \includegraphics[height=3.2cm, width=8.5cm]{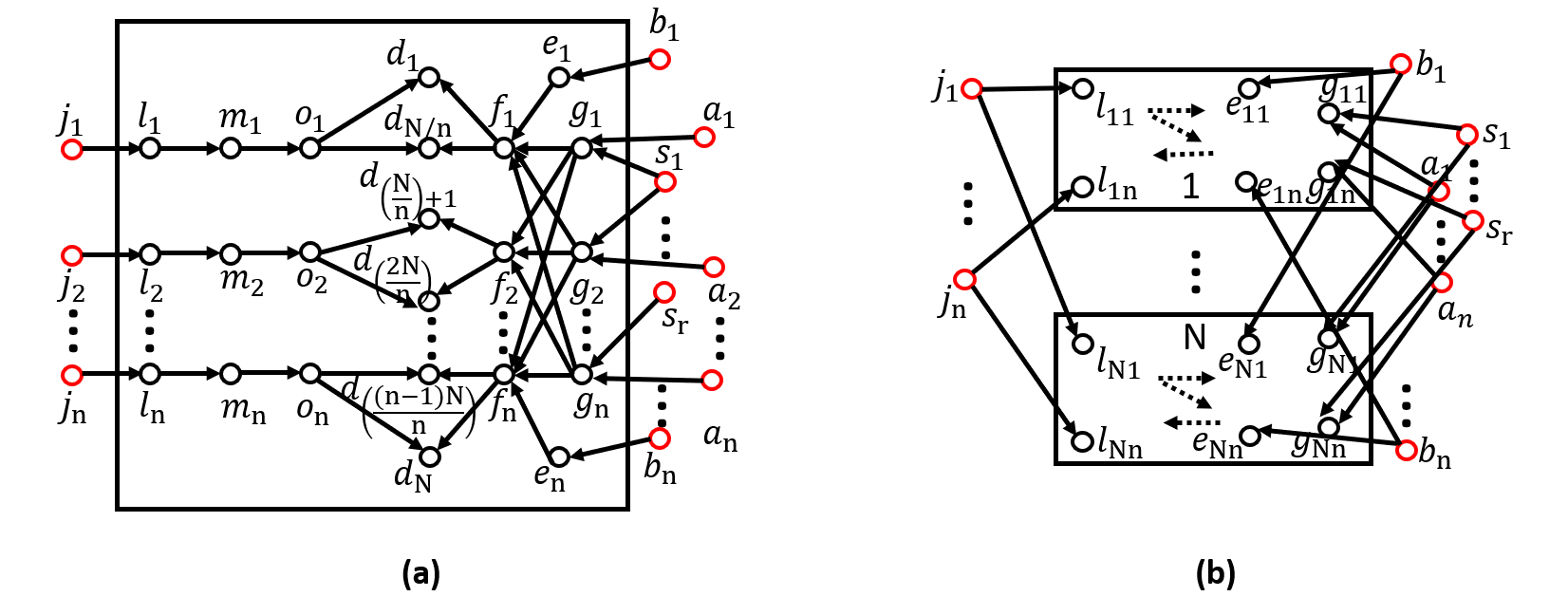}
    \caption{Social network: (a) The structure of one copy, $\calJ'$;  (b) Instance $\calJ$, obtained from $N$ copies of the structure shown on the left side; seeds of $i_2$: $\{a_1, ..., a_n\}$; seeds of $i_3$: $\{b_1, ..., b_n\}$; seeds of $i_4$: $\{j_1, ..., j_n\}$. }
    \label{fig:inapprox}
\end{figure}

We can set the values and prices of items and itemsets such that $\util(\{\2,\3\}) < c/4\cdot \util(\{\1,\4\})$ (see Fig. \ref{fig:util_tables} (b)). Choosing a sufficiently large $N$: $N > max\{k/c, 8n/c\}$, we can see that each term in the expression $(**)$ above is strictly less than $cN^2\util(\{i1,i4\})$. There are 4 terms in the expression $(**)$ and their sum is $< 4\times c/4 \times N^2 \times \util(\{\1,\4\})$. Thus, the optimal welfare on a NO-instance is 
\begin{equation}\label{eq:wn}
x_n^* = (*) < 4 \times c/4 \times N^2 \times \util(\{\1,\4\}) = c\times N^2 \times \util(\{\1,\4\}). 
\end{equation}

Hence combining this with Claim \ref{claim2} we get,
\eat{ 
YES instance:

For all the the copies we have the same adoption behavior. $n$ number of $g$ nodes and $n$ number of $f$ nodes adopt $i_1$. $n$ number of $e$ nodes adopt $i_3$. $3n$ number of $l,m$ and $o$ nodes adopt $i_4$. $N$ number of $d$ nodes adopt $\{i_1, i_4\}$. Thus the welfare is $$N(2n \util(i_1) + n \util(i_3) + 3n \util(i_4) + N \util(\{i_1, i_4\}))$$

In addition the seeds contribute the following. $r$ number of $s$ nodes adopt $i_1$, $n$ number of $a$ nodes adopt $i_2$, $n$ number of $b$ nodes adopt $i_3$ and $n$ number of $j$ nodes adopt $i_4$. Thus the total welfare,
$$N(2n \util(i_1) + n \util(i_3) + 3n \util(i_4) + N \util(\{i_1, i_4\}))+$$
$$r \util(i_1) + n (\util(i_2) + \util(i_3) + \util(i_4))$$
}

%Given a SET COVER instance $\inst$, transform it into an instance $\calJ$ of \prob and run the algorithm $\calA$ on $\calJ$. We have:
\vspace*{-1ex} 
\begin{claim}\label{claim3}
Given a SET COVER instance $\inst$, transform it into an instance $\calJ$ of \prob and run the algorithm $\calA$ on $\calJ$.
$\inst$ is a YES-instance iff the welfare returned by $\calA$ is $>c N^2\util(\{\1,\4\})$. $\inst$ is a NO-instance iff the welfare returned by $\calA$ is $<c N^2\util(\{\1,\4\})$
\end{claim}

\begin{proof} 

Suppose $\inst$ is a YES-instance. Then, by Claim~\ref{claim2}, the optimal welfare of $\calJ$ is $> N^2\util(\{\1,\4\})$, so the welfare returned by Algorithm $\calA$ on $\calJ$ is $> c N^2\util(\{\1,\4\})$. Suppose $\inst$ is a NO-instance. Then the optimal welfare of $\calJ$ is $x^*_n < c N^2\util(\{\1,\4\})$. Thus, even if Algorithm $\calA$ returned the optimal welfare on the NO-instance $\calJ$, it would be strictly less than the welfare returned on the corresponding YES-instance.

For a NO-instance the optimal welfare is upper bounded by Eq. \ref{eq:wn}. Hence, the claim follows. 

%%%% TO RESURRECT FOR FULL VERSION 
\end{proof}
 
\vspace*{-1.5ex} 
The theorem follows, $\calA$ cannot exist unless $P = NP$.\qed 
\end{proof}

%%%%%%%%%%%%%%%%%%% 
\eat{ 
Ignoring the smaller terms we have,
\begin{align*}
x^*_n &\leq kN\util(\4) + (N-k)N\util(\{\2,\3\}) \\
&< kN\util(\4) + (N-k)Nc\util(\4) \;\; \text{(since $\util(\{\2,\3\}) < c\util(\4)$}, \\
& \text{\hspace*{32ex} see Fig. \ref{fig:util_tables} (b)}\\
&=  N^2c\util(\4) + Nk(1-c)\util(\4) \\
&\approx cN^2\util(\4) \text{, assuming $k<<N$}
\end{align*}

\note[Laks]{$\approx$ doesn't work in this reduction. The issue is that the optimal welfare for a no instance could be higher than $cN^2\util(\4)$, so the gap is blurred.}
}
%%%%%%%%%%%%%%%%%%%

%%%% TO RESURRECT FOR FULL VERSION 

\begin{center}
    
\begin{table}[]
\begin{tabular}{|l|l|l|l|}
\hline
Item bundle          & Value & Price & Utility \\ \hline
$\emptyset$          & 0     & 0     & 0       \\ \hline
$\1$                   &  15.1  &  10    &  5.1     \\ \hline
$\2$                   & 105   &  100   & 5       \\ \hline
$\3$                   &  105   & 100   &  5       \\ \hline
$\4$                   & 101   &  1     &  100      \\ \hline
$\1,\2$                 & 114.9 & 110   & 4.9     \\ \hline
$\1,\3$                 & 114.9 & 110   & 4.9     \\ \hline
$\1,\4$                 & 116.1 & 11    & 105.1   \\ \hline
$\2,\3$                 & 210   & 200   & 10      \\ \hline
$\2,\4$                 & 206   & 101   & 105     \\ \hline
$\3,\4$                 & 206   & 101   & 105     \\ \hline
$\1,\2,\3$               & 214.6 & 210   & 4.6     \\ \hline
$\1,\2,\4$               & 214   & 111   & 103     \\ \hline
$\1,\3,\4$               & 214   & 111   & 103     \\ \hline
$\2,\3,\4$               & 210.5   & 201   & 9.5       \\ \hline
$\1,\2,\3,\4$             & 214.6 & 211   & 3.6     \\ \hline
\end{tabular}
\caption{Utility configuration for different item bundles}
\label{tab:utility-table}
\end{table}

\end{center}
%%%% TO RESURRECT FOR FULL VERSION 

\eat{ 
Now for a contradiction suppose there is a $PTIME$ $c$-approximation algorithm $A$, for \prob. Let $w$ the welfare that $A$ outputs. Clearly $cx^*\leq w \leq x^*$, where $x^*$ is the optimal welfare. Since for YES-instance $x^*_y > N^2 \util(\4)$, $w>c N^2  \util(\4)$. On the other hand, for NO-instance, $x^*_n \leq c N^2  \util(\4) $, hence $ w \leq c N^2 \util(\4)$. Since these two intervals donot, if there is such $A$, we can use it to solve the decision version of set cover in $PTIME$. This cannot be true unless $P=NP$. 

\end{proof}
}

\vspace{-2mm} 
\section{Approximation Algorithms}\label{sec:algo}
%%%%%%%%%%%%%%%%%%%%%%%%%%%%%%%%%%%%%%%%%%%%
%%%%%%%%%%%%%%%%%%%%%%%%%%%%%%%%%%%%%%%%%%%%
\eat{
\weic{
Things to do:
\begin{itemize}
	\item 
	explain the marginal version of $\PRIMM^+$.
	
	\item 
	probably remove the explanation on pure competition, since we do not special results for it.
	
	\item 
	other changed explanations.
	
	\item 
	Fix the proofs for MaxGRD, removing the pure competition part. 
	
	\item 
	add the summary of the results of the two algorithms in the end, to rpove the main Theorem~\ref{thm:main}.
\end{itemize}
}
}
%%%%%%%%%%%%%%%%%%%%%%%%%%%%%%%%%%%%%%%%%%%%%%%%%%
%%%%%%%%%%%%%%%%%%%%%%%%%%%%%%%%%%%%%%%%%%%%%%%%%%

Since the \prob problem cannot be approximated within any constant factor in general, in this section we propose 
	several approximation algorithms that
	either produce a non-constant approximation guarantee dependent on the problem instance or
	a constant approximation guarantee for a special case of \prob.
We first define some important notions.

\spara{Truncated utility}
For accounting the social welfare of an allocation, we develop the notion of \textit{truncated utility} of an item. Recall that when the noise of an item makes its utility negative, no node adopts the item. Hence what contributes to the final expected social welfare is %\sout{the part of the noise that makes the utility positive.} 
the set of non-negative contributions to utility.  
We call this  the \emph{truncated utility}, denoted  $\utilt(I) := max(0, \util(I))$. Thus for a (node, item) allocation pair $(v, i)$, its expected social welfare
(when there are no other allocations) is $\rho(v,i) = \mathbb{E}[\utilt(i)] \sigma(\{v\})$, where $\sigma(\{v\})$ is the influence spread of $\{v\}$.

\spara{Minimum and maximum utility bundle}
We define $\umin = \min_{i\in \allitems} \E[\cU^+(i)]$ as the minimum expected truncated utility of any item in $\allitems$, and 
$\umax = \E[\max_{I \subseteq \allitems} \cU^+(I) ]$ as the expected maximum truncated utility of any item {\em bundle} in $\allitems$. 
Note that the definitions of $\umin$ and $\umax$ are not  symmetric: (a) $\umin$ takes the minimum of an expectation, while
	$\umax$ takes the expectation of a maximum; and (b) $\umin$ takes minimum on single items while $\umax$ takes maximum among all  bundles.
The reason of this asymmetry will be clear in our analysis.

\spara{Superior and inferior item}
A given itemset $\allitems$ is said to have a \emph{superior} item $i_m$, if the least possible utility of $i_m$  is 
strictly higher than the highest possible utility of any item in $\allitems \setminus \{i_m\}$. 
\eat{\pink{If there are multiple such items, we arbitrarily choose one of them as the superior item.}} Notice the definition of superior item entails that the noise distribution should be bounded in some way. We discuss a practical way to bound the noise in our experiments (\textsection \ref{sec:exp}). Given a superior item, all the other items of the itemset are called \emph{inferior} items. 
\eat{
\note[Laks]{Can items be tied on their max possible utility? A strict interpretation of the def. does not allow ties for superior items - \red{addressed, see the pink line}} 
}   

In what follows, we present three different algorithms with progressively better theoretical guarantees,  under progressively stronger assumptions. 
As a preview, our first algorithm \sgrd\ provides a $\frac{\umin}{\umax}(1-\frac{1}{e})$-approximation in the most general case. 
Our second algorithm, \mgrd, assumes no prior allocations, i.e.,   $\allalloc^p = \emptyset$. 
Under this assumption, it provides a  $\frac{1}{m}(1-\frac{1}{e})$-approximation, where $m$ is the number of items. By simply returning the better of the two allocations produced by \sgrd and \mgrd, the bound is improved to $max\{\frac{\umin}{\umax},\frac{1}{m}\}(1-\frac{1}{e})$, when $\allalloc^p = \emptyset$. Our final algorithm \supgrd\ assumes that there exists a superior item in the itemset, the allocations for all  inferior items are fixed, and that items exhibit pure competition. Under these assumptions, it provides a $(1-\frac{1}{e})$-approximation.

\label{sec:msgrd}
\subsection{SeqGrd Algorithm}
\eat{ 
We first present algorithm \sgrd and show that under the general setting,  %of \model, 
\sgrd achieves a  $\frac{\umin}{\umax}(1 - \frac{1}{e})$ approximation. } 
The pseudocode of our first algorithm \sgrd is shown in Algorithm~\ref{alg:sgrd}.

\begin{algorithm}
\caption{$\sgrd(G, \epsilon, \ell, \allallocp, I_2, \bvec)$} \label{alg:sgrd}
\begin{algorithmic}[1]
\State $S^P \leftarrow$ Seed nodes of the allocation $\allallocp$ 
%\State $\bmax \leftarrow \sum_$
\State $\seqSeeds \leftarrow  \PRIMAP(G, \epsilon, \ell, S^P, \bvec, \sum_{i \in I_2} b_i)$  \label{lin:sgrd_PRIMM} 
\State $\seqAlloc \leftarrow \emptyset$ 
\State Sort $I_2$ in decreasing order of the expected truncated utility \label{lin:sgrdp_sort}
%\State remove all nodes in $\seqSeeds$ that appear in $\allallocp$
\State $\Added \leftarrow \emptyset$
\For{$i \in I_2$}
\State $\seqSeeds_i \leftarrow \text{ top } b_i \text{ nodes from } \seqSeeds$ \label{lin:msgrd_assign2}
\If{$\rho(\seqAlloc \mid \allallocp) < \rho((\seqAlloc \cup (\seqSeeds_i \times \{i\})) \mid \allallocp)$} \label{lin:msgrd_positives}
\State $\seqAlloc \leftarrow \seqAlloc \cup (\seqSeeds_i \times \{i\})$
\State Remove those $b_i$ nodes from $\seqSeeds$
\State $\Added \leftarrow \Added \cup \{i\} $
\EndIf \label{lin:msgrd_positivee}
\EndFor	
\For{$i \in I_2 \setminus \Added $}\label{lin:msgrd_randoms}
\State $\seqSeeds_i \leftarrow \text{ top } b_i \text{ nodes from } \seqSeeds$ 
\State $\seqAlloc \leftarrow \seqAlloc \cup (\seqSeeds_i \times \{i\})$
\State Remove those $b_i$ nodes from $\seqSeeds$
\EndFor	\label{lin:msgrd_randomse}
\State Return $\seqAlloc$
\end{algorithmic}
\end{algorithm}

Algorithm \sgrd\ considers the general setting where a set of items have already been seeded and 
$\allallocp$ corresponds to this partial allocation. Let $S^P := \{v \mid (v,i)\in \allallocp\}$  be the seed set allocated in $\allallocp$ and  let $I_2$ denote the remaining items which 
have yet to be allocated. The algorithm takes  a graph $G$, to-be-allocated itemset $I_2$, 
	item budget vector $\bvec$ for the items in $I_2$, accuracy parameter $\epsilon$, tolerance parameter $\ell$, the partial allocation $\allallocp$ as input. %and item budget vector $\bvec$ 
	It first selects a seedset $\seqSeeds$ of size $\overline{b}$, where $\overline{b} := \sum_{i \in I_2} b_i$ \footnote{We assume  $I_1 \cap I_2 = \emptyset$.} 
	\eat{If $I_1\cap\ I_2\ne\emptyset$, $\overline{b} := \sum_{i \in I_1 \cup I_2} b_i$.
	\weic{Why do we want to consider $I_1 \cap I_2 \ne \emptyset$? I feel that we should only consider the case that $I_1 \cap I_2 = \emptyset$.}} 
	(line \ref{lin:sgrd_PRIMM}). To select the seeds it uses an algorithm, called $\PRIMAP$, which delivers a set of seeds that are approximately optimal w.r.t. the marginal gain $\sigma(S | S^P)$. \eat{and has a property called \textit{prefix preservation on the marginals}.} We \eat{explain this property formally and} present the $\PRIMAP$ algorithm in \textsection \ref{sec:algod} and establish its properties. 
	
\sgrd then sorts the items based on their truncated utility (line \ref{lin:sgrdp_sort}). Starting from the item $i$ having the highest truncated utility, it tries to allocate the item to the top $b_i$ nodes of $\seqSeeds$, $\seqSeeds_i$.
If the allocation $\seqSeeds_i \times \{i\}$ yields a positive marginal welfare, it is added to the existing allocation and nodes of $\seqSeeds_i$ are removed for future considerations (Lines \ref{lin:msgrd_positives}-\ref{lin:msgrd_positivee}).
The items that are not allocated in this iteration are appended following an arbitrary order (lines \ref{lin:msgrd_randoms}-\ref{lin:msgrd_randomse}) and allocated at the end. 
\eat{
\note[Laks]{Random or arbitrary order? \red{arbitrary. addressed}} We now prove the approximation bound of this algorithm.
} 

%Recall from the definitions in Section \ref{sec:msgrd} that $\umin$ is the minimum expected truncated utility of any item in $\allitems$, while $\umax$ is the  expected value of the maximum truncated utility of any item bundle in $\allitems$. 
Let $\Gamma_w(S)$ be the set of nodes reachable from a seed set $S$ in the possible world $w$.
We first establish  the following lemma.
\begin{lemma} \label{lem:condexpinq}
	Let $\allalloc$ be an allocation, $S$ be its seedset, 
	let $w$ be a random possible world. Then
	for any node $v\in V$, we have
	\[\umin \le \E_w \left[ \cU_w(\cA^{\allalloc}_w(v)) \mid v\in \Gamma_w(S)\right] \le \umax.\]
\end{lemma}
\begin{proof}
	Let $w=(w_1, w_2)$ where $w_1$ is the edge possible world and $w_2$ is the noise possible world.
	Note that, (a) reachable set $\Gamma_w(S)$ is only determined by the edge possible world $w_1$, so we can use $\Gamma_{w_1}(S)$ to represent it;
	(b) utility function $\cU_w(\cdot)$ is only determined by the noise possible world $w_2$, so we can use $\cU_{w_2}(\cdot)$ to represent it; and
	(c) adoption set $\cA^{\allalloc}_w(v)$ is determined by both $w_1$ and $w_2$, so we use $\cA^{\allalloc}_{w_1,w_2}(v)$ to represent it.
	\begin{align}
	& \E_w \left[ \cU_w(\cA^{\allalloc}_w(v)) \mid v\in \Gamma_w(S)\right]  \nonumber \\
%	& = \E_{w_1} \left[ \E_{w_2} \left[ \cU_w(\cA^{\allalloc}_w(v)) \mid v\in \Gamma_w(S), w_1 \right] \right]  \nonumber \\
	& = \E_{w_1} \left[ \E_{w_2} \left[ \cU_{w_2}(\cA^{\allalloc}_{w_1,w_2}(v)) \mid v\in \Gamma_{w_1}(S)\right] \right]. \label{eq:condexp}
	\end{align}
	We first prove the lower bound $\umin$.
	To do so, we prove that for any fixed edge possible world $w_1$ and conditioned on $v \in \Gamma_{w_1}(S)$, we have 
	$\E_{w_2} \left[ \cU_{w_2}(\cA^{\allalloc}_{w_1,w_2}(v))\right] \ge \umin = \min_i \E[\cU^+(i)] = \min_i \E_{w_2}[\cU^+_{w_2}(i)] $.
	Once this is proved, from Eq.~\eqref{eq:condexp}, we immediately have 
	$\E_w \left[ \cU_w(\cA^{\allalloc}_w(v)) \mid v\in \Gamma_w(S)\right] \ge \E_{w_1}[ \umin ] = \umin$.
	
	Consider first a seed $u\in S$.
	Let $\cA^{\allalloc}_{w_1,w_2}(u,1)$ be the set of items adopted by $u$ initially at time $1$ before the propagation starts. 
	Let $I^u$ be the set of items allocated to $u$ in $\allalloc$. 
	Note that $I^u$ is determined purely by the fixed allocation $\allalloc$ and is not affected by the noise or edge possible world.
	By our model, node $u$ will select the best item bundle in $I^u$ and adopt them as $\cA^{\allalloc}_{w_1,w_2}(u,1)$. 
	Then we know that $\cU_{w_2}(\cA^{\allalloc}_{w_1,w_2}(u,1)) \ge \max_{i\in I^u} \cU_{w_2}^+(i) $.
	Therefore, we have
	\begin{align*}
	& \E_{w_2} \left[ \cU_{w_2}(\cA^{\allalloc}_{w_1,w_2}(u,1)) \right] \ge \E_{w_2} \left[ \max_{i\in I^u} \cU_{w_2}^+(i) \right] \\
	& \ge  \max_{i\in I^u}  \E_{w_2} \left[ \cU_{w_2}^+(i) \right] \ge \min_{i\in \allitems} \E_{w_2} \left[ \cU_{w_2}^+(i) \right] = \umin.
	\end{align*}

\eat{ 	
	\note[Laks]{The above derivation leaves me wondering if we could tighten the lower bound: e.g., we relax from $\max_{i\in I^u}  \E_{w_2} \left[ \cU_{w_2}^+(i) \right]$ to $\min_{i\in \allitems} \E_{w_2} \left[ \cU_{w_2}^+(i) \right]$. \red{Not Addressed}}
	
	\weic{For Laks' comment above, perhaps we cannot tighten the bound, because $I^u$ is an arbitrary allocation on node $u$, and it could only contain the minimum utility item.}
	} 
	
	This means that for the initial seed adoption, we have that their expected utility is at least $\umin$.
	Now for any $v \in \Gamma_{w_1}(S)$, $v$ is reachable from some seed node $u \in S$ via some shortest path in the edge possible world $w_1$.
\eat{ 
	\note[Laks]{There are multiple errors below: firstly time begins at $t=1$ not $0$ and secondly $\cA$ takes time as an argument not as a subscript, based on earlier definitions.\red{Addressed}} 
	} 
	By the propagation model, then the utility of $v$'s final adoption $\cU_{w_2}(\cA^{\allalloc}_{w_1,w_2}(v))$ should be at least the utility of $u$'s initial 
	adoption, $\cU_{w_2}(\cA^{\allalloc}_{w_1,w_2}(u,1))$.
	Then we have $\E_{w_2} \left[ \cU_{w_2}(\cA^{\allalloc}_{w_1,w_2}(v))\right] \ge \E_{w_2} \left[ \cU_{w_2}(\cA^{\allalloc}_{w_1,w_2}(u,1)) \right] \ge \umin$.
	This concludes the proof.
	
	The proof of the upper bound $\umax$  is straightforward:
	$\E_w \left[ \cU_w(\cA^{\allalloc}_w(v)) \mid v\in \Gamma_w(S)\right] \le \E_w \left[ \max_{I\subseteq \allitems} \cU^+(I) \mid v\in \Gamma_w(S)\right]
	= \E_w \left[ \max_{I\subseteq \allitems} \cU^+(I)\right] = \umax$. \qed 
\end{proof}

\begin{lemma} \label{lem:lowerupper}
Let $\allalloc$ be an allocation and $S$ its corresponding seed nodes. Then $\umin \cdot \sigma(S) \le \rho(\allalloc) \le \umax \cdot \sigma(S)$. 
	
\end{lemma}
\eat{ 
\note[Laks]{We used $S^{\allalloc}$ previously to denote the seeds of an allocation. But we could perhaps 
use that opportunity to simplify notation. See Sec. 3 (pink), e.g. \red{Addressed}} 
} 
\begin{proof}
The lower bound is derived below:
\begin{align*}
&\rho(\allalloc) = \E_w\left[\sum_v \cU_w(\cA^{\allalloc}_w(v)) \right] = \sum_v \E_w \left[ \cU_w(\cA^{\allalloc}_w(v)) \right]\\
& = \sum_v \Pr_w[v\in \Gamma_w(S) ] \cdot \E_w \left[ \cU_w(\cA^{\allalloc}_w(v)) \mid v\in \Gamma_w(S)\right] \\
& \ge \sum_v \Pr_w[v\in \Gamma_w(S) ] \cdot \umin  = \umin \cdot \sigma(S),
\end{align*}
where the inequality is by Lemma~\ref{lem:condexpinq}.
The upper bound can be shown in a similar way. \qed 
\end{proof}

We are now ready to prove the following bound for $\sgrd$. 

\begin{theorem}\label{lem:sgrd_bound}
Let $\seqAlloc$ be the allocation returned by the Algorithm \sgrd. Given $\epsilon, \ell >0$, we have $\sw(\seqAlloc \cup \allallocp) \geq \frac{\umin}{\umax}(1-\frac{1}{e}- \epsilon)\sw(\allalloc^A \cup \allallocp)$ w.p. at least $1 - \frac{1}{|V|^\ell}$, where $\allalloc^A$ is any arbitrary allocation of items in $I_2$
	respecting the budget constraint.
\end{theorem}
\begin{proof}
Let $S^{Seq}$, $S^A$ and $S^P$ be the seed sets of the allocations $\seqAlloc$, $\allalloc^A$ and $\allallocp$ respectively.
Then $|S^A| \le \sum_{i\in I_2} b_i$.
By \sgrd, $S^{Seq}$ exhausts all budgets for items in $I_2$, 
\eat{thus
 the total number of seeds in the allocation $\allalloc^{Seq}$ is} so  $|S^{Seq}| = \sum_{i\in I_2} b_i$.
Since $S^{Seq}$ are the top seeds returned by $\PRIMM^+$, we have that w.p. at least $1 - \frac{1}{|V|^\ell}$, 
$$\sigma(S^{Seq} \mid S^P) \geq (1 - \frac{1}{e} - \epsilon) \sigma(S^A \mid S^P).$$
From this, it follows that 
$$\sigma(S^{Seq} \cup  S^P) \geq (1 - \frac{1}{e} - \epsilon) \sigma(S^A \cup S^P).$$
Therefore, we have
\begin{align*}
&\rho(\allalloc^{Seq} \cup \allallocp) \ge \umin \cdot \sigma(S^{Seq} \cup S^P) \\
& \ge \umin \cdot (1-1/e - \epsilon) \cdot \sigma(S^A \cup S^P) \\
&\ge \frac{\umin}{\umax}(1-\frac{1}{e}- \epsilon) \rho(\allalloc^A \cup \allallocp),
\end{align*}
where the first and the last inequality follow from Lemma~\ref{lem:lowerupper}, while the 
middle inequality follows from using $\PRIMAP$.  
\end{proof}

We note that the property of $\PRIMAP$ that is exploited in the proof above is its ability to select seed nodes $S$ 
such that they are approximately optimal w.r.t. the marginal gain over an existing seed set $S^P$. The prefix 
preserving on marginals property of $\PRIMAP$ is not needed in the above proof. However, our next algorithm \mgrd relies on 
	the prefix-preserving property.

\hspace*{2ex} 

\noindent
{\bf  \sgrdi\ \ Algorithm}

The proof of the approximation bound above does not rely on marginal check (Algorithm~\ref{alg:sgrd}, line \ref{lin:msgrd_positives}).  \eat{Thus, the approximation guarantee of \sgrd\  holds without the marginal check. } 
	\eat{Thus even when we do not perform the marginal check, the same approximation holds.} 
	We call the  version of \sgrd that does not perform marginal check  \sgrdi (No Marginal). 
	Specifically, \sgrdi simply sorts the items based on their truncated utility, allocates item $i$ to the first $b_i$ nodes of $S^{Grd}$, where $S^{Grd}$ is selected using $\PRIMAP$, and removes those $b_i$ nodes from $S^{Grd}$.

Computing  marginals involves sampling, which takes significant time in large networks. On the other hand, the marginal check avoids the phenomenon of items with lower (truncated) utility blocking those with higher utility,  to some extent. Thus even though \sgrdi is  faster than \sgrd and has the same approximation guarantee, under certain utility configurations, \eat{when avoiding this sort of item blocking is essential,} the welfare  produced by  \sgrdi  can be worse than that of \sgrd. We explore this  in our experiments in \textsection{\ref{sec:exp}}.
On the other hand, we still append all items in the end to exhaust the budget in \sgrd (lines~\ref{lin:msgrd_randoms}--\ref{lin:msgrd_randomse}).
To really discard a certain itemset, we need to exhaustively search through all itemset combinations, which is time-consuming. So we only do a simple marginal check in \sgrd, and append all items at the end to ensure the theoretical guarantee.

\subsection{MaxGrd Algorithm}

Our next algorithm \mgrd provides  $\frac{1}{m}(1-\frac{1}{e})$-approximation, when $\allallocp = \emptyset$, i.e., no prior allocation.
\eat{ 
\note[Laks]{The pseudocode doesn't match this description. There is no marginal there. 
Did you want to only use \mgrd when there is no previous allocation? In this case, use of $I_2$ w/o 
any explanation is confusing. 
In general, there is a confusing array of notation used in the paper which may annoy a reviewer: e.g., $S, S_i, S^{\allalloc}, S^{Grd}$, 
to mention a few. I am not sure if all of these are necessary. Look for ways to consolidate notation. \red{Addressed}} 
} 
The pseudocode is shown in Algorithm \ref{alg:mgrd}. Like \sgrd, \mgrd also selects its seedset $\maxSeeds$ using $\PRIMAP$, but the size of the seedset  is different: $\bmax := \max_{i \in I_2} b_i$, i.e., the maximum budget of any unallocated item (line \ref{lin:mgrd_PRIMM}). Then for every item $i \in I_2$, it computes the expected marginal social welfare of the allocation $\rho( (\maxSeeds_i \times \{i\}) \mid \allallocp)$, where $\maxSeeds_i$ is the set of first $b_i$ nodes of $\maxSeeds$. It returns the allocation with the maximum welfare (line \ref{lin:mgrd}). 

\begin{algorithm}
\caption{$\mgrd(G, \epsilon, \ell, \allallocp, I_2, \bvec)$} \label{alg:mgrd}
\begin{algorithmic}[1] 
\State $\maxSeeds \leftarrow  \PRIMAP(G, \epsilon, \ell, S^P, \bvec, \max_{i \in I_2} b_i)$  \label{lin:mgrd_PRIMM} 
\State $\maxSeeds_i \leftarrow \text{top $b_i$ nodes of $\maxSeeds$}$, $\forall i\in I_2$
    \State $i_{max} \leftarrow \argmax_{i \in I_2}\{ \rho(\maxSeeds_i \times \{i\} \mid \allallocp)  \}$\label{lin:mgrd}
    \State Return $\maxSeeds_{i_{max}} \times \{ i_{max} \}$
\end{algorithmic}
\end{algorithm}

{Notice that \mgrd is applicable even when $S^p \neq \emptyset$, so we have provided the algorithm for this general case. However, it enjoys an approximation bound only for the special case, when $S^p=\emptyset$. } 
We prove the following lemma under this constraint, which is instrumental in the proof of the approximation bound. A key observation is that given a possible world $w$, the utility function $\util_w(\cdot)$ in that possible world is submodular. This follows from the fact that valuation is submodular and price and noise, being additive are both modular. 
\vspace{-2mm}
\eat{ 
\begin{lemma}\label{lem:bound3}
In any arbitrary allocation $\allalloc^A$, let $S_i$ be the seed nodes of item $i$, then when $\allallocp = \emptyset$\eat{or under pure competition}, $\rho(\cup_{i=1}^m (S_i \times \{i\}) \cup \allallocp ) \leq \sum_{i=1}^m \rho( (S_i \times \{i\}) \cup \allallocp)$.
\end{lemma}
} 

\begin{lemma}\label{lem:bound3}
Let  $\allalloc := \cup_{i=1}^m (S_i \times \{i\})$ be an arbitrary allocation, where $S_i$ is the set of seed nodes of item $i$. Then  $\rho(\cup_{i=1}^m (S_i \times \{i\})) \leq \sum_{i=1}^m \rho( (S_i \times \{i\}))$.
\end{lemma} 

\begin{proof}[Sketch] 
Consider an arbitrary but fixed possible world $w$ and an arbitrary item $i\in I_2$.
Let $v$ be any node that adopts $i$ in  $w$ under the allocation $\cup_{i=1}^m (S_i \times \{i\})$. 
We can show %below 
that $v$ must also adopt $i$ in $w$ when the allocation is only $(S_i \times \{i\})$. The lemma follows from this. 
\end{proof} 
\vspace{-2mm}
\begin{theorem}\label{lem:mgrd_bound}
	Suppose that $\allallocp = \emptyset$.
Let $\maxAlloc$ be the allocation produced by \mgrd. Given $\epsilon, \ell >0$,  we have $\sw(\maxAlloc) \geq \frac{1}{m}(1-\frac{1}{e}- \epsilon)\sw(\allalloc^A)$ w.p. at least $1 - \frac{1}{|V|^\ell}$, where $\allalloc^A$ is any arbitrary allocation.
\end{theorem}

\begin{proof}
Recall that item $i$ has a budget $b_i$ and expected utility $u_i$. 
Since in an arbitrary allocation $|S^A_i| \leq b_i$, from the prefix preserving property of $\PRIMAP$  we have,
\begin{equation} \label{eq:2Soft}
\sigma(\maxSeeds_i) \geq \left(1 - \frac{1}{e} - \epsilon \right) \sigma(S^A_i).
\end{equation}
Let $\mathbb{E}[\util^+(i)]$ be the expected positive utility of item $i$.
We have $\rho(\maxSeeds_i \times \{i\}) = \mathbb{E}[\util^+(i)]\cdot \sigma(\maxSeeds_i) $ and
	$\rho(S^A_i \times \{i\}) = \mathbb{E}[\util^+(i)]\cdot \sigma(S^A_i)$.
Therefore, from Eq.\eqref{eq:2Soft} we have
\begin{equation} \label{eq:welfare2pure}
\rho(\maxSeeds_i \times \{i\}) \geq \left(1 - \frac{1}{e} - \epsilon \right) \rho(S^A_i \times \{i\}) .
\end{equation}
When $\allallocp = \emptyset$, using Eq. \ref{eq:welfare2pure} and Lemma \ref{lem:bound3}, we have 
\begin{align*}
&\rho(\maxAlloc \cup \allallocp) = \rho(\maxAlloc) = \max_{i \in I_2}\{\rho(\maxSeeds_i \times \{i\})\}		\\
&\ge  \frac{1}{m} \sum_{i=1}^m \rho(\maxSeeds_i \times \{i\})\geq \frac{1}{m} \left(1-\frac{1}{e}- \epsilon \right) \sum_{i=1}^m \rho(S^A_i \times \{i\} )\\
&\geq \frac{1}{m} \left(1-\frac{1}{e}- \epsilon \right)  \rho(\cup_{i=1}^m (S^A_i \times \{i\}) )= \frac{1}{m} \left(1-\frac{1}{e}- \epsilon \right) \rho(\allalloc^A).
\end{align*}
\end{proof}

\textbf{Can \mgrd produce better welfare than \sgrd?} 
{Hypothetically, there can be situations where \mgrd can produce better welfare than \sgrd.  E.g., consider a network with nodes $\{u, v, w, x\}$ and edges $\{(u,v), (v,w), (x,w)\}$ where all edge probabilities are $1$. There are two items $i, j$, with  all noise terms being $0$. The utilities are $\util(\{i\}) = 10, \util(\{j\})=1, \util(\{i,j\})=0$ and both items $i$ and $j$ have a budget of $1$. Then \sgrd\ will yield the allocation $\allalloc^{Seq} = \{(u,i), (x,j)\}$, resulting in a social welfare of $2\times 10 + 1\times 2 = 22$. On the other hand, \mgrd\ will only allocate $u$ to $i$, resulting in a social welfare of $3\times 10 = 30$. } 

{In our experiments, however, we find that situations where \mgrd\ dominates \sgrd\ are rare. We hypothesize that this is because in a large network, with a number of seeds that is a small fraction of the network size $n$, blocking caused by the allocation of seeds to additional items by \sgrd\ is less likely to occur. } 
\eat{The highest utility item is $i_1$ (utility $10$). Both $i_2$ and $i_3$ (each with utility $1$) exhibit pure competition against $i_1$, however between themselves they exhibit soft competition: $\util(\{i_2,i_3\}) = 1.5$

Hence throwing away $i_2$ and $i_3$ altogether is not ideal. Thus in \sgrd, every allocation of the $i_2$ and $i_3$ blocks the propagation of $i_1$ and hence produces negative marginal. \sgrd will still allocate $i_2$ and $i_3$ (Line 14), resulting in a drop in the welfare from the allocation of those. However, \mgrd will allocate only one item, $i_1$. Consequently, the negative marginal producing allocation will be avoided. Such a situation is rare in real networks because the budget is typically much less than the network size, hence it is not possible to reach all the nodes by allocating one item. Hence in our experiments (\textsection \ref{sec:exp}) we always find \sgrd to produce better welfare.}  

{Note that the approximation guarantee of  \sgrd  holds also when $\allalloc^p = \emptyset$. Thus running both \sgrd and \mgrd individually and returning the allocation with higher welfare 
would achieve a $max\{\frac{\umin}{\umax},\frac{1}{m}\}(1-\frac{1}{e})$-approximation, as a consequence of
 Theorems \ref{lem:sgrd_bound} and \ref{lem:mgrd_bound}.}

\subsubsection{$\PRIMAP$}\label{sec:algod}

We now present our $\PRIMAP$ algorithm used by \sgrd and \mgrd to select seeds. 
First, we formally present the property of prefix preservation on marginals.

\eat{ 
\note[Laks]{Make sure notations are used consistently, e.g., the right symbol, font, etc. \red{Addressed}}
} 

\begin{definition}{\sc (Prefix Preservation on Marginals).\ } \label{def:PRIMM}
	Given $G$ $=$ $(V,E,p)$, budget vector $\bvec$, the number of seeds to be selected $\bmax$ and a fixed seed set $S^P$, an influence maximization algorithm $\mathbb{A}$
	is prefix-preserving on marginals w.r.t. $\bvec$ and $S^P$, 
	if for any $\epsilon$ $>$ $0$ and $\ell$ $>$ $0$, 
	$\mathbb{A}$  returns an ordered set $S$ of size $\bmax$, 
	such that   w.p.  at least $1$ $-$ $\frac{1}{|V|^{\ell}}$, $\sigma(S \mid S^P)$ $\geq$ $(1-\frac{1}{e} - \epsilon)$ $\OPT_{\bmax \mid S^P}$ and
	for every $b_i$ $\in$ $\bvec$, the first $b_i$ nodes of $S$, denoted $S_{i}$, satisfies $\sigma(S_{i} \mid S^P)$ $\geq$ $(1-\frac{1}{e} - \epsilon)$ $\OPT_{b_i \mid S^P} $, %\sigma(\optSeeds_{b_i})$, 
	where %$\optSeeds_{b_i}$ is the optimal seed set with $b_i$ seeds in the IC model.
	$\OPT_{b \mid S^P}$ is the optimal marginal expected spread of $b$ nodes on top the existing seeds $S^P$.
\end{definition}

\eat{ \note[Laks]{Saying that the top-$b_i$ nodes need to satisfy something means you always get to use the TOP $b_i$ nodes as opposed to the NEXT $b_i$ nodes after previous seeds are committed. \red{Addressed}} } 

In \cite{ban2019}, the authors proposed a seed selection algorithm called $\PRIMM$ that is prefix-preserving in spread, using the Reverse Reachable Sets (RR-sets), as proposed in IMM \cite{tang15}. Here, we  modify the standard RR-set construction slightly to account for the presence of existing seed set $S^P$: Given an existing allocation $\allalloc^P$, we construct a marginal RR-set as follows. Choose a root node $v\in V$ uniformly at random, add it to $R_v$ and start a BFS from $v$. Whenever $u\in R_v$, sample each incoming edge $(u',u)$
w.p. $p_{u'u}$ and add it to $R_v$. Stop when no new nodes are added to $R_v$; if at any stage $R_v$ overlaps $S^P$, i.e., if $R_v\cap S^P\ne\emptyset$, then set $R_v := \emptyset$. That is, whenever a generated RR-set ``hits" $S^P$, just set it to $\emptyset$. Algorithm \ref{alg:marsample} shows the pseudo code of this marginal RR-set sampling process. Given graph $G$, a number $\theta$ denoting how many RR-sets needs to be sampled and a fixed seed nodes $S^P$, $Marginal\_Sampling$ generates $\theta$ number of RR-sets to $\mathcal{R}$ from $G$, based on the marginal on $S^P$.

\begin{algorithm}
\caption{$Marginal\_Sampling(G, \mathcal{R}, \theta, S^P)$} \label{alg:marsample}
\begin{algorithmic}[1] 
\While {$|\mathcal{R}| \leq \theta$}
    \State Select $v$ from $G$ uniformly at random
    \State $R \leftarrow BFS(v)$
    \If {$R \cap S^P \neq \emptyset$}
        \State $R \leftarrow \emptyset$
    \EndIf
    \State $\mathcal{R} \leftarrow \mathcal{R} \cup R$
\EndWhile
\State Return $\mathcal{R}$
\end{algorithmic}
\end{algorithm}

$\PRIMAP$ using $Marginal\_Sampling$, achieves the property of prefix preservation on marginals. Its pseudo code is shown in Algorithm \ref{alg:primap}.

\begin{algorithm}
\caption{$\PRIMM$ $( G, \epsilon, \ell, S^P, \bvec, \bmax )$} \label{alg:primap}
\begin{algorithmic}
\State Initialize $\mathcal{R} = \emptyset$, $s = 1$, $n = |V|$, $i = 1$, $\epsilon^\prime = \sqrt{2} \cdot \epsilon$, $\budgetSwitch = \false, \bvec = \bvec \cup \bmax$
\State $\ell = \ell  + \log 2 / \log n $, $\ell^\prime = \log_n(n^{\ell} \cdot |\bvec|)$ \label{lin:ellb}
\While {$i \leq \log_2(n) - 1$ and $s \leq |\bvec|$}
	\State $k = b_s$, $LB =1$
	\State $x = \frac{n}{2^i}$; $\theta_i = \lambda^\prime_k/x$, where $\lambda^\prime_k$ is defined in Eq. \eqref{eq:lambdap}
	\State $Marginal\_Sampling(G,\mathcal{R}, \theta_i, S^P)$
	\If {$\budgetSwitch$}
		\State $S_k = $ the first $k$ nodes in the ordered set $S_{b_{s-1}}$ returned from
		the previous call to $\nodeselect$
	\Else
		\State $S_k =\nodeselect(\mathcal{R}, k)$
	\EndIf
	\If {$n \cdot F_{\mathcal{R}}(S_k) \geq (1 + \epsilon^\prime)\cdot x$} \label{lin:cov}
		\State $LB = n \cdot F_{\mathcal{R}(S_k)} / (1+\epsilon^\prime) $
		\State $\theta_k = \lambda^\ast_k / LB$, where $\lambda^\ast_k$ is defined in Eq. \eqref{eq:lambdaa} \label{lin:suc1}
		\State $Marginal\_Sampling(G,\mathcal{R}, \theta_k, S^P)$
		\State $s = s + 1 $; $\budgetSwitch = \true$	
	\Else 
	\State $i= i+1$; $\budgetSwitch = \false$ \label{lin:fail}
	\EndIf
	
\EndWhile
\If {$s \leq |\bvec|$}
	\State $\theta_k = \lambda^\ast_{b_s}/LB$ \label{lin:final}	
\EndIf	
\State $\mathcal{R} = \emptyset$
\State $Marginal\_Sampling(G,\mathcal{R}, \theta_k, S^P)$
\State $S_{\bmax} = \nodeselect(\mathcal{R}, \bmax)$ \label{lin:nodes}
\textbf{return} $S_{\bmax}$ as the final seed set;
\end{algorithmic}
\end{algorithm}

It runs in time $O((\bmax+\ell + \log_n |\bvec|)(n+m)\log \ n \cdot \epsilon^{-2})$, where $\bmax := \sum_{i \in I_2} b_i$ for \sgrd and $\bmax := \max_{i \in I_2} b_i$, for \mgrd. 

\subsection{SupGrd Algorithm} \label{sec:alg_supgrd}
Our third algorithm \supgrd  provides a constant $(1-\frac{1}{e})$-approximation. The bound holds under more restrictive conditions as given below. 

\spara{Conditions required for \supgrd approximation bound}
(i) There exists a superior item (defined in \textsection \ref{sec:msgrd}) $i_m$ in the item set: i.e., under any noise possible world $w_2$, $\util_{w_2} (i_m)> \util_{w_2}(i), \forall i \in \allitems \setminus \{i_m\}$. (ii) Seeds for all the inferior items are fixed: that is, $I_2 = \{i_m\}$ is the only item for which an allocation needs to be found; and (iii) There is pure competition between all items: every node can adopt at most one item.
Under these conditions, we first show that the social welfare is monotone and submodular. 

\begin{lemma} \label{lem:superiormonotone}
Given $\allallocp$ and $I_2$, let $\allalloc_1$ and $\allalloc_2$ be two allocations over $I_2$ such that $\allalloc_1 \subseteq \allalloc_2$. Then $\rho(\allalloc_1 \cup \allallocp) \leq \rho(\allalloc_2 \cup \allallocp)$.
\end{lemma}

\begin{proof}
In an arbitrary but fixed possible world $w= (w_1, w_2)$, $\util_{w_2}( \adopt^{\allalloc_1 \cup \allallocp}_{w_1, w_2}(v)) \leq \util_{w_2}( \adopt^{\allalloc_2 \cup \allallocp}_{w_1, w_2}(v))$. This is because if $v$ changes its adoption between the two allocations, then it must change it to $i_m$ since all other inferior item seeds are fixed. Since $i_m$ is the superior item, the claim holds.
\eat{ 
Now,
\begin{align*}
\rho_{w = (w_1,w_2)}(\allalloc_1 \cup \allallocp) &= \sum_{v \in V} \util_{w_2}( \adopt^{\allalloc_1 \cup \allallocp}_{w_1, w_2}(v)) \\
&\leq \sum_{v \in V} \util_{w_2}( \adopt^{\allalloc_2 \cup \allallocp}_{w_1, w_2}(v)) \\
&= \rho_{w = (w_1,w_2)}(\allalloc_2 \cup \allallocp)
\end{align*}
}
Since this holds for every $w$, the lemma follows.
\end{proof}

\begin{lemma}
Given $\allallocp$ and $I_2$, let $\allalloc_1$ and $\allalloc_2$ be two allocations over $I_2$ such that $\allalloc_1 \subseteq \allalloc_2$. Let $s = (u, i_m) \notin \allalloc_2$ be an allocation pair. Then $\rho(s \mid \allalloc_1 \cup \allallocp) \geq \rho(s \mid \allalloc_2 \cup \allallocp)$.
\end{lemma}

\begin{proof} 
Let $w = (w_1, w_2)$ be a arbitrary but fixed possible world. Let $C_i$ denote the set of all nodes that adopt $i_m$ under allocation $s$ but \emph{not} under $\allalloc_i \cup \allallocp$, $i = 1, 2$. Then $C_2 \subseteq C_1$. Thus,
\begin{align*}
\rho_{w_1,w_2}(s \mid \allalloc_2 \cup \allallocp) &= \sum_{v \in C_2} (\util_{w_2}(\{i_m\}) - \util_{w_2}( \adopt^{\allalloc_2 \cup \allallocp}_{w_1, w_2}(v))) \\
&\leq \sum_{v \in C_1} (\util_{w_2}(\{i_m\}) - \util_{w_2}( \adopt^{\allalloc_1 \cup \allallocp}_{w_1, w_2}(v)))\\ 
%&  \\ %\text{,using monotonicity}\\
& = \rho_{w_1,w_2}(s \mid \allalloc_1 \cup \allallocp)
\end{align*}
\eat{ 
\begin{align*}
\rho_{w_1,w_2}(s \mid \allalloc_2 \cup \allallocp) &= \sum_{v \in C} (\util_{w_2}(\{i_m\}) - \util_{w_2}( \adopt^{\allalloc_2 \cup \allallocp}_{w_1, w_2}(v))) \\
&\geq \sum_{v \in C} (\util_{w_2}(\{i_m\}) - \util_{w_2}( \adopt^{\allalloc_1 \cup \allallocp}_{w_1, w_2}(v)))\\ 
&  \text{,using monotonicity}\\
&\geq \rho_{w_1,w_2}(s \mid \allalloc_1 \cup \allallocp)
\end{align*}
} 

Since this holds for every $w$, the lemma follows.
\end{proof}

Since social welfare is monotone and submodular, a standard greedy selection based on the marginal welfare will have $(1 - \frac{1}{e})$-approximation. However since computing spread itself is \#P-hard, computing the exact marginal is not feasible. In IM, sampling using RR-sets has been used to achieve state of the art performance. In what follows, by extending IMM \cite{tang15}, we adopt a martingale approach for seed selection in \supgrd. Given $\epsilon$ and $\ell$,  \supgrd returns a seed set that has a $(1-\frac{1}{e} - \epsilon)$-approximation w.p. at least $1 - \frac{1}{n^\ell}$.

In the classical setting, RR-set samples are used to compute an unbiased estimation of the spread. In our case we need to estimate the \emph{marginal welfare} using the RR-sets. 
Towards that we define a notion of weight for every RR-set. The weight of an RR-set $R_v$ denotes the marginal gain in the expected social welfare achieved by activating the root $v$ of the RR-set $R_v$. 
Thus it is the difference between the expected truncated utility of the item that the root $v$ adopts under the existing partial allocation $\allallocp$ and that of $i_m$. 
To ensure that the root $v$ indeed adopts $i_m$, the path from some seed of $i_m$ to $v$ should be no longer than that from any seed of $\allallocp$ to $v$. Thus, a weighted RR-set is constructed as follows. 

\begin{definition} \label{def:rrset}
(Weighted Reverse Reachable Set). For a given fixed allocation $\allallocp$ and a
node $v \in G$, a weighted RR-set of $v$, $R_v$ is obtained by starting with $R_v=\{v\}$ and starting a BFS from $v$ such that: for $u\in R_v$, sample each incoming edge $(u',u)$
w.p. $p_{u',u}$ and add it to $R_v$; stop when either no new nodes are added or $R_v$ overlaps $S^P$ 
(so the distance from any node in $R_v$ to $v$ along the reversely generated edges is 
	at most the distance from $S^P$ to $v$).
Then, the weight of $R_v$ is $w(R_v) = \utilt(\{i_m\}) -  max_{i \in I^s \mid s \in S^P \cap R(v)} \utilt(i)$, where $I^s$ denotes the items allocated to node $s$ in \eat{$S^P$, seeds of} the allocation $\allallocp$.
\end{definition}

\supgrd samples RR-sets using an early termination as described in Definition \ref{def:rrset}. This construction ensures that if any member of a {weighted} RR-set is seeded with $i_m$, the root of the RR-set, $v$, will adopt $i_m$. In what follows, we first  establish the connection between marginal social welfare and weighted RR-sets and then present efficient seed selection and RR-set sampling algorithms to maximize the marginal social welfare. \eat{When the context is clear, we drop $S^P$ from the notation for simplicity.} 

 For a node set $S$, let $\mathbb{I}[.]$ be an indicator function denoting whether $S$ covers the (weighted) RR set $R$, i.e., $\mathbb{I}(S \cap R \ne \emptyset) = 1$, if $S \cap R_v \neq \emptyset$, $0$ otherwise. Also let $\mathcal{L}(G)$ denote the distribution of all the live edge graphs, then extending the result of Borg et al., we now prove the following lemma for  weighted RR-sets.

\begin{lemma} \label{lem:borgs}
For given seed sets $S$ and $S^P$, we have
$\mathbb{E}_{w_1 \sim G}[\rho_{w_1}(S \mid S^P)] = n \cdot \mathbb{E}_{ v \sim V, w_1 \sim G} [\mathbb{I}(S \cap R_v \ne \emptyset) \cdot w(R_v)]$
where $n = |V|$ is the number of nodes in $G$.
\end{lemma}

\begin{proof}

\begin{align*}
\mathbb{E}_{w_1 \sim \mathcal{L}(G)}\rho_{w_1}(S \mid S^P)
&= \mathbb{E}_{w_1 \sim \mathcal{L}(G)} \left[\sum_{v \in V} \mathbb{I}(S \cap R_v \ne \emptyset) \cdot w(R_v) \right] \\
&= n \cdot \mathbb{E}_{ v \sim V, w_1 \sim \mathcal{L}(G)} [ \mathbb{I}(S \cap R_v = 1) \cdot w(R_v)]
\end{align*}
\end{proof}

We now extend the RR-set based efficient approximation IM algorithm, IMM, for maximizing welfare. Similar to IMM, our algorithm $\supgrd$ has two key phases, namely, $NodeSelection$ and $Sampling$. The $NodeSelection$ phase is similar to that of IMM, except we consider the weight of  RR-sets while selecting  seed nodes.  \eat{using $coverage$.} For a node set $S$ and  a collection of weighted RR-sets $\mathcal{R}$, define 
$M_{\mathcal{R}}(S) := \sum_{R \in \mathcal{R}} \mathbb{I}[S \cap R \ne \emptyset] \cdot w(R)$. Let $b' = b_{i_m}$ be the budget of the superior item $i_m$. Given a set $\mathcal{R}$, $NodeSelection$ selects $b'$ seeds that maximizes $M_{\mathcal{R}}$ (Algorithm \ref{alg:NS}). 

\begin{algorithm}[t!]

\caption{$NodeSelection(\mathcal{R}, b')$} \label{alg:NS}
\begin{algorithmic}[1]
\State Initialize $S_{b'} = \emptyset, i=1$
\While{$i \leq b'$}
	\State Select $v \in V \setminus S_{b'}$ which has the highest $M_{\mathcal{R}}(S_{b'} \cup v) - M_{\mathcal{R}}(S_{b'})$
	\State $S_{b'} \leftarrow S_{b'} \cup \{v\}$
	\State Remove $R$ from $\mathcal{R}$ if $v \in R$ 
	%$\mathbb{I}(v \in R) = 1$ 
\EndWhile
\State \textbf{return} $S_{b'}$ as the final seed set
\end{algorithmic}
\end{algorithm}

Next, the goal of the $Sampling$ phase (Algorithm \ref{alg:sam}) is to generate $\mathcal{R}$ such that $|\mathcal{R}| \geq \lambda /OPT$, where $OPT$ is the optimal welfare, and $\lambda$ is defined as follows,
\begin{equation}\label{eq:lambdaa}
\lambda = 2n \cdot ((1 - 1/e)\cdot \alpha + \beta)^2 \cdot \epsilon^{-2}, 
\end{equation} 
where, $\alpha = \sqrt{\ell \log n + \log 2}$ and \\ 
$\beta = \sqrt{(1 - 1/e) \cdot (\log \tbinom{n}{b'}+\ell \log \ n + \log 2)}$.

Since $OPT$ is unknown, the $Sampling$ first ensures that it finds a lower bound to $OPT$ w.h.p. For that it deploys a statistical test using a binary search on the range of $OPT$. The maximum possible value of the welfare $OPT$ is $UB = n \times \umax$, when every node in the network adopts the superior item $i_m$, $\umax$ is the utility of $i_m$. Thus the binary search ranges from $1$ to $UB$ (Line \ref{lin:loop}).

\begin{algorithm}[t!]
\caption{$Sampling(G, k, \epsilon, \ell)$} \label{alg:sam}
\begin{algorithmic}[1]
\State Initialize $\mathcal{R} = \emptyset$, $LB = 1$, $UB = |V| \times \umax$, $i = 1$, $\epsilon^\prime = \sqrt{2} \cdot \epsilon$, $\ell = \ell + log \ 2/log \ n$. \label{lin:init}
\For{$i = 1$ to $log_2 UB -1$ } \label{lin:loop}
	\State $x = n / 2^i$, $\theta_i = \lambda' / x$
	\While{$|\mathcal{R}| \leq \theta_i$}
		\State Add a random RR set to $\mathcal{R}$
	\EndWhile
	\State $S_i = NodeSelection(\mathcal{R}, k)$
	\If{$\frac{n}{\theta} \cdot M_{\mathcal{R}}(S_i) \geq (1 + \epsilon') \cdot x$} \label{lin:lb}
		\State $LB = \frac{n}{\theta} \cdot M_{\mathcal{R}} / (1 + \epsilon') $
		\State Break;
	\EndIf
\EndFor
\State $\mathcal{R} = \emptyset$ \label{lin:rrfinal}
\While{$|\mathcal{R}| \leq \lambda / LB$}
	\State Add a random RR set to $\mathcal{R}$
\EndWhile
\end{algorithmic}
\end{algorithm}

A good lower bound is found when the condition of Line \ref{lin:lb} is satisfied. Different from \cite{tang15}, this condition directly operates on welfare. This is a key step in the correctness of the algorithm, hence we prove it explicitly in Lemma \ref{lem:lb}.

\begin{lemma} \label{lem:lb}
Let $x \in [1, UB], \epsilon'$ and $\delta \in (0,1)$, then if we invoke $NodeSelection$ with $|\mathcal{R}|=\theta$, where 
\begin{equation}
\theta \geq \frac{(2+ \frac{2}{3} \epsilon') \cdot (log \tbinom{n}{b'} + log (1/\delta))}{\epsilon'^2} \cdot \frac{n}{x}
\end{equation}
and $S$ is the output $NodeSelection$ returns,
then if $OPT <x$, $\frac{n}{\theta} \cdot M_{\mathcal{R}}(S) < (1 + \epsilon') \cdot x $, w.p. at least $(1 - \delta)$. 
\end{lemma}

\begin{proof}
Let $x_i$ be a random variable for each $R_i \in \mathcal{R}$ defined as, $x_i = \frac{w(R_i) \cdot \mathbb{I}(S \cap R_i \neq \emptyset)}{w_{max}}$, where $w_{max}$ is the maximum weight possible for any RR set. Thus, $0 \leq x_i \leq 1$, which ensures the martingle property.
Now let $F_{\mathcal{R}}(S) = \frac{M_{\mathcal{R}}(S)}{w_{max}}$, $p = \mathbb{E}[F_{\mathcal{R}}(S)]$ and $\alpha = \frac{(1+\epsilon')\cdot x}{np \cdot w_{max}} - 1$, Using Lemma \ref{lem:borgs} and linearity of expectation,

\begin{align*}
p&= \mathbb{E}[F_{\mathcal{R}}(S)] = \mathbb{E}[\frac{M_{\mathcal{R}}(S)}{w_{max} }] = \mathbb{E}[\rho(S \mid S^p)]/(w_{max} ) \\
&\leq OPT /(w_{max}  \cdot n) \leq x/(w_{max} \cdot n)
\end{align*}

Consequently $\alpha > \epsilon' \cdot x/(np) $ and from Lemma 6 of \cite{tang15},

\begin{align*}
&Pr[\frac{n}{\theta} \cdot M_{\mathcal{R}}(S) \geq (1+\epsilon') \cdot x] \leq \delta/ \tbinom{n}{b'} 
\end{align*}
Finally by applying union bound we get $\frac{n}{\theta} \cdot M_{\mathcal{R}}(S) < (1 + \epsilon') \cdot x $, w.p. at least $(1 - \delta)$.
\end{proof}

Thus by setting $\lambda'$ using Eq. \eqref{eq:lambdap}, we get Theorem 2 of \cite{tang15}

\begin{equation}\label{eq:lambdap}
\lambda^\prime = \frac{(2+\frac{2}{3}\epsilon^\prime) \cdot (\log {n \choose b'} + \ell^\prime \cdot \log \ n+\log\log_{2} \ n )\cdot n}{\epsilon^{\prime 2}},
\end{equation}

The rest of the proof is similar to that of \cite{tang15}, which gives us the following result.

\begin{theorem}\label{lem:supgrd_bound}
Let $\allallocp$ be a partial allocation on the inferior items.
Let $\greedAlloc$ be the allocation of the superior item produced by {\rm SupGrd}. Given $\epsilon, \ell >0$, we have $\sw(\greedAlloc \cup \allallocp) \geq (1-\frac{1}{e}- \epsilon)\sw(\allalloc^A \cup \allallocp)$ w.p. at least $1 - \frac{1}{|V|^\ell}$, where $\allalloc^A$ is any arbitrary allocation.
\end{theorem}

\eat{
\begin{proof}
From Theorem 1 of \cite{tang15}, $NodeSelection$ returns a seed set  which provides a $(1-1/e-\epsilon)$-approximate marginal welfare  w.p.  at least $1-1/n^\ell$. However, following the fix proposed in \cite{chen2018issue}, we need to generate all the RR-sets afresh as done in Line \ref{lin:rrfinal}. Then from Lemma \ref{lem:lb}, and  union bound, by setting $\ell = \ell + log \ 2/log \ n$ in line \ref{lin:init}, we bound the failure probability to at most $\frac{1}{|V|^\ell}$. The Lemma follows. 
\end{proof}
}

\noindent
{\bf Running time: }
Let $w_{min}$ be the minumum weight of an RR set. Then
using Lemma $9$ of \cite{tang15}, the expected total time to generate $\mathcal{R}$ is determined by,
\begin{align*}\label{eq:rtime}
\mathbb{E}[\sum_{R \in \mathcal{R}}wid(R)] &= \mathbb{E}[|\mathcal{R}|] \cdot EPT \nonumber \\
& \leq O((b'+\ell)n\log \ n \cdot \epsilon^{-2})/OPT \cdot \frac{m}{n} OPT/w_{min} \\
&= O((b'+\ell)(n+m)\log \ n \cdot \epsilon^{-2}/w_{min})
\end{align*}

Notice that generating an RR-set from scratch for the final node selection (line \ref{lin:rrfinal}), following the fix of \cite{chen2018issue}, only adds a multiplicative factor of $2$. Hence the overall asymptotic running time to generate $\mathcal{R}$ remains unaffected.

\vspace*{-1ex} 
\section{Experiments}\label{sec:exp}
\begin{table}[t!]
	\scriptsize
	\begin{tabular}{r|c|c|c|c|c|}
	 & \net & \dbBook & \dbMovie & \orkut & \twit \\ \hline
	{ \# nodes}			& $15.2$K & $23.3$K	& $34.9$K & $3.07$M & $41.7$M \\ 
	{ \# edges} 		& $31.4$K  & $141$K	& $274$K & $117$M & $1.47$G \\ 
	 { avg. deg.}     & $4.13$ & $6.5$ & $7.9$ & $77.5$ & $70.5$\\ 
	 { type}            &  undirected  & directed & directed  &  undirected & directed  \\ \hline
	\end{tabular}
	\caption{Network Statistics}
	\vspace{-2mm}
	 \label{tab:datasets}
\end{table}

%%%%%%%%%%%%%%%%%%%%%%%%%%%%%%%%%%%%%%%%%%%%%%%%%%%%%%%%%%%%%%%%%%%%
%%%%% All the Fig.s here

\begin{figure*}[ht]
\begin{small}
\hspace*{-12mm}\includegraphics[height=2.55cm,width=1.15\textwidth]{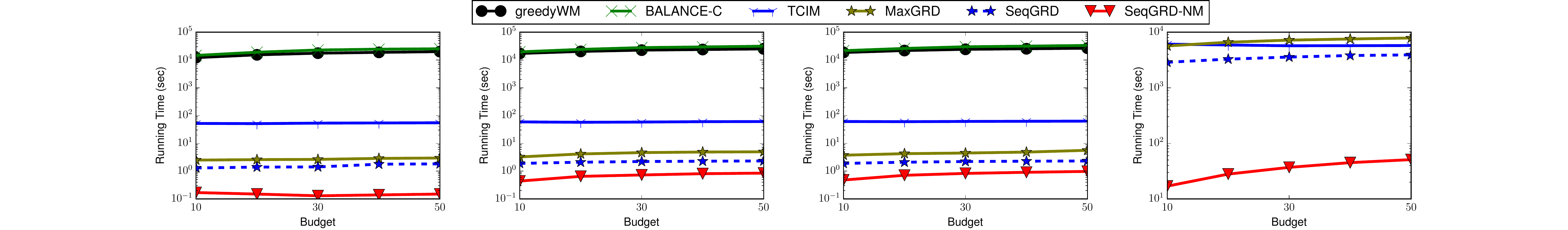}
\begin{tabular}{cccc}
%\hspace{-6mm}\includegraphics[width=.25\textwidth]{figs/pdf/flxt.pdf} &
%\hspace{-2mm}\includegraphics[width=.25\textwidth]{figs/pdf/dbt.pdf} &
%\hspace{-2mm}\includegraphics[width=.25\textwidth]{figs/pdf/dmt.pdf} &
%\hspace{-2mm}\includegraphics[width=.25\textwidth]{figs/pdf/twt.pdf} \\
\hspace{2mm} (a)  NetHept  \hspace{24mm} &  (b) \dbBook \hspace{21mm} &  (c) \dbMovie \hspace{21mm} &  (d)  \orkut
\end{tabular}
\caption{Running times of $\gm$, $\bc$, $\tc$, $\mgrd$, $\sgrd$ and $\sgrdi$ (on Configuration $1$)} \label{fig:time}
\vspace{-2mm}
\end{small}
\end{figure*}

\begin{figure*}[ht]
\begin{small}
\vspace{-2mm}
\hspace*{-12mm}\includegraphics[height=2.55cm,width=1.15\textwidth]{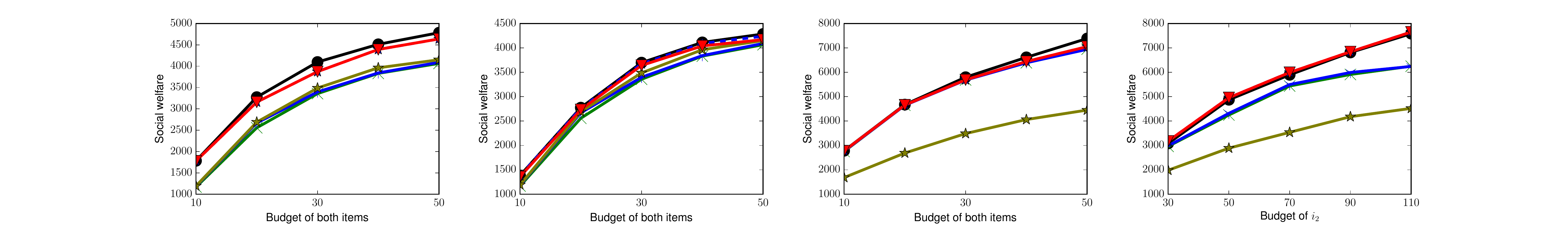} %[height=3.5cm,width=1.22\textwidth]
\begin{tabular}{cccc}
%\hspace{-6mm}\includegraphics[width=.25\textwidth]{figs/pdf/flxt.pdf} &
%\hspace{-2mm}\includegraphics[width=.25\textwidth]{figs/pdf/dbt.pdf} &
%\hspace{-2mm}\includegraphics[width=.25\textwidth]{figs/pdf/dmt.pdf} &
%\hspace{-2mm}\includegraphics[width=.25\textwidth]{figs/pdf/twt.pdf} \\
 \hspace{8mm} (a) Configuration $1$  \hspace{17mm} &  (b) Configuration $2$ \hspace{17mm} &  (c) Configuration $3$ \hspace{17mm} &  (d)  Configuration $4$
\end{tabular}
\caption{Expected social welfare in four configurations (on the \dbMovie network)} \label{fig:welfare}
\vspace{-2mm}
\end{small}
\end{figure*}

\begin{figure*}[ht]
\begin{small}
\vspace{-2mm}
\hspace*{-12mm}\includegraphics[height=2.55cm,width=1.15\textwidth]{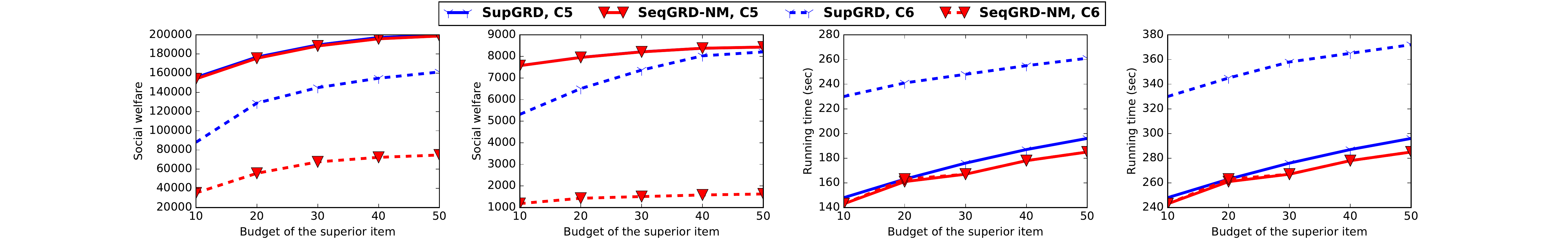}
\begin{tabular}{cccc}
(a) \orkut & \hspace{23mm}  (b) \twit \hspace{27mm} &  (c) \orkut \hspace{27mm} &  (d)  \twit
\end{tabular}
\vspace{-2mm}
\caption{Comparison between SupGRD and SeqGRD on C2 and C3 (a-b) Social welfare, (c-d) Running time} \label{fig:exp_supgrd}
%\vspace{-4mm}
\end{small}
\end{figure*}

\begin{figure*}[ht]
\begin{small}
\vspace{-2mm}
\hspace*{-12mm}\includegraphics[height=2.55cm,width=1.15\textwidth]{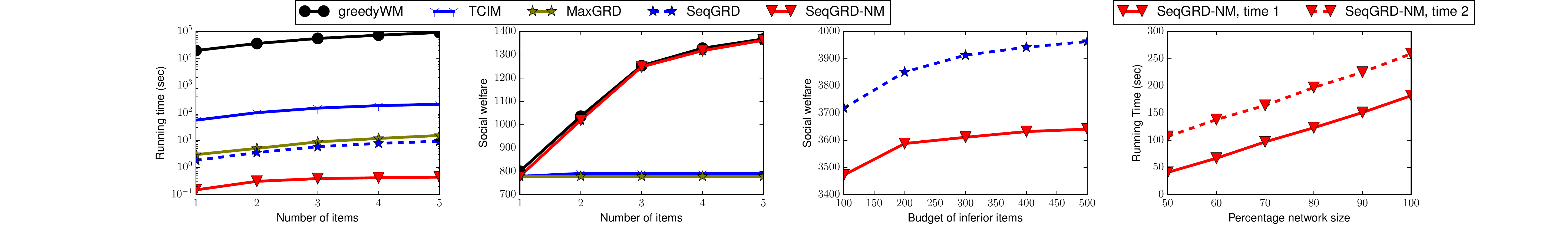} %[height=3.5cm,width=1.22\textwidth]
\begin{tabular}{cccc}
%\hspace{-6mm}\includegraphics[width=.25\textwidth]{figs/pdf/flxt.pdf} &
%\hspace{-2mm}\includegraphics[width=.25\textwidth]{figs/pdf/dbt.pdf} &
%\hspace{-2mm}\includegraphics[width=.25\textwidth]{figs/pdf/dmt.pdf} &
%\hspace{-2mm}\includegraphics[width=.25\textwidth]{figs/pdf/twt.pdf} \\
 \hspace{8mm} (a) Running Time \hspace{17mm} &  (b) Social Welfare \hspace{17mm} &  (c) Social welfare \hspace{17mm} &  (d)  Running time
\end{tabular}
%\vspace{-2mm}
\caption{Multi-item experiments: Impact of number of items on (a) Running time, (b) Social welfare on NetHept. (c) Comparing performance of \sgrd and \sgrdi on NetHept. (d) Scalability on \orkut} \label{fig:assorted}
\vspace{-2mm}
\end{small}
\end{figure*}

\begin{figure*}[ht]
\begin{small}
\hspace*{-12mm}\includegraphics[height=2.55cm,width=1.15\textwidth]{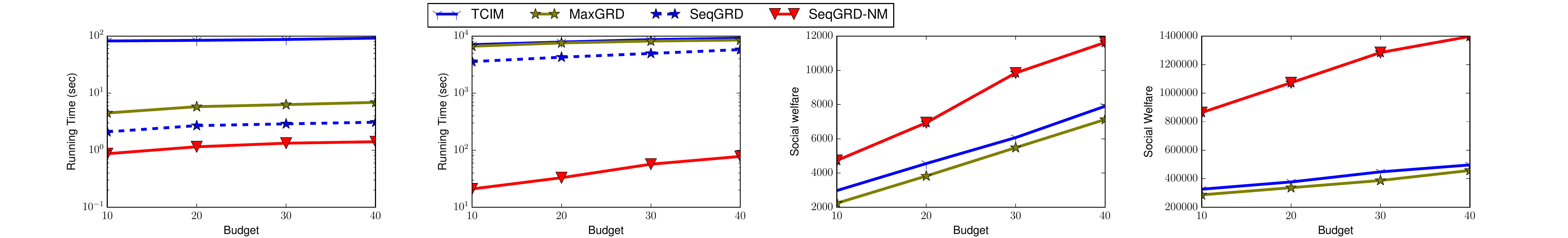}
\begin{tabular}{cccc}
%\hspace{-6mm}\includegraphics[width=.25\textwidth]{figs/pdf/flxt.pdf} &
%\hspace{-2mm}\includegraphics[width=.25\textwidth]{figs/pdf/dbt.pdf} &
%\hspace{-2mm}\includegraphics[width=.25\textwidth]{figs/pdf/dmt.pdf} &
%\hspace{-2mm}\includegraphics[width=.25\textwidth]{figs/pdf/twt.pdf} \\
\hspace{-5mm} (a)  NetHept  \hspace{34mm} &  (b) Orkut \hspace{33mm} &  (c) Nethept \hspace{31mm} &  (d)  \orkut
\end{tabular}
\caption{Performance of $\tc$, $\mgrd$, $\sgrd$ and $\sgrdi$ on real utility configurations (Table \ref{tab:real_utils})} \label{fig:real_exp}
\vspace{-2mm}
\end{small}
\end{figure*}

%%%%%%%%%%%%%%%%%%%%%%%%%%%%%%%%%%%%%%%%%%%%%%%%%%%%%%%%%%%%%%%%%%%%%%%%%%%%%%%%%%%%%%%%%%%
%\vspace{-4mm}

\subsection{Experiment Setup}

All our experiments are run on a Linux machine with Intel Xeon $2.6$ GHz CPU and $128$ GB RAM.

\spara{6.1.1 Networks} 
Our experiments were conducted on five real social networks: \net, \db, \dm, \twit, and \orkut, whose characteristics are summarized in~Table \ref{tab:datasets}. Of these, \net, \db, and \dm are benchmarks in IM literature \cite{lu2015arxiv}, while \twit and \orkut are two of the largest public networks available at \cite{twitter}.

\spara{6.1.2 Algorithms compared}
In the experiments our four algorithms -- \sgrd, \sgrdi, \mgrd, and \supgrd are compared against three baselines --  \tc, \bc and \gm. 
There is no previous work that can deal with both arbitrary degree of competition and multiple items in propagation.
Our first two baselines each covers one aspect. 
\tc \cite{lin2015analyzing} in particular assumes a propagation model which is an extension of the IC model under pure competition. It can, however, handle more than two items. Given fixed seed sets of other competing items, \tc selects seeds of an item under a budget constraint, such that the number of adoptions of that item is maximized. When we run \tc for multiple items, we select seeds for each of the items one by one, while keeping the seeds of other items fixed and then report the allocation that produces the maximum welfare.

In contrast, \bc \cite{garimella2017balancing} does not assume pure competition, but it works only when number of items in propagation is two. Given an initial seed placement of the two items, \bc chooses the remaining seeds such that at the end of the propagation, the number of nodes seeing either both the items or none, is maximized. Thus for competing ideas, \bc ensures that there is a balanced exposure of the two ideas to the most number of nodes. It is non-trivial to extend \bc for more than two items hence we compare against it only in two item set up. 

Both {\tc} and \bc aim to maximize adoption count, not social welfare. Our third baseline \gm maximizes the social welfare directly. It greedily selects iteratively the (node, item) pair that maximizes the marginal social welfare, till the budgets are exhausted. Below, by deterministic utility of an itemset $I$, we mean $\val(I) - \price(I)$, i.e., its utility with the noise term ignored.

\spara{6.1.3 Default parameters}
Following previous works \cite{Huang2017,Nguyen2016} we set probability of edge $e = (u,v)$ to $1/\ind(v)$, where $\ind(v)$ is the in-degree of node $v$.
Unless otherwise specified, we use $\epsilon = 0.5$ and $\ell = 1$ as our default in all the algorithms that use these parameters. 
We test the algorithms across a wide variety of utility configurations to cover different aspects of competition. We will describe the configurations as we present the corresponding experiments. Whenever marginal gains are required, we run 5000 simulations and take the average result. 
%\note[Laks]{Check the above.} 
\vspace{-2mm}
\subsection{Experiments with two items}

For our first set of experiments we restrict the number of items to two so that we can compare against all of the mentioned baselines. We also consider four different configurations to capture different kinds of competition. The details of the configurations are given in Table \ref{tab:configs}. 
In configurations C1 and C2, the items exhibit pure competition. In C1, items have comparable individual utility.
In C2, the difference between individual utility is high:  $i$'s deterministic utility is $1$, 10 times higher than that of $j$. 
%\weic{Please check with the table. It seems C1 is the one with camparable utility and C2 is the one with high difference in utility.}
C3 and C4 exhibit soft competition. 
Except for C4, in all  configurations we consider the same budget for both items (uniform); budget is varied from 10 to 50 in steps of 10. In C4, we fix the budget of $i$ to 50 and vary $j$'s budget (non-uniform) from 30 to 100 in steps of 20. We assume $\allalloc^p = \emptyset$ in these configurations. Since it does not meet constraints required by \supgrd, we defer the comparison until \textsection 6.2.3.
%\weic{In Table~\ref{tab:configs}, the noise column, it shows $\{i, j\}: N(0,2)$. Why so? Isn't that the noise of $i$ and $j$ are independent? If so, do not need
%the noise for $i$ and $j$ together. They are just the sum of two independent noise variables. If written as $\{i, j\}: N(0,2)$, this suggests a new variable for
%the noise.}

\spara{6.2.1 Running time}
First we compare the running time of the algorithms using C1 as a representative case. Fig. \ref{fig:time} shows the result on four networks. \sgrdi is orders of magnitude faster than other algorithms in every network. The reason is that \sgrdi does not compute any marginal. 
Each marginal computation requires iterating over 5000 samples, which significantly increases the running time. For the same reason \gm and \bc exhibit exorbitantly high running time: they do not in fact complete in 6 hours on a large network like \orkut. Hence they are not included in Fig. \ref{fig:time}(d). Except for \sgrdi, none of the other algorithms scale to the largest network \twit. We will compare \sgrdi and \supgrd on \twit later. Performance on other configurations show similar trends, and hence omitted for brevity.

\begin{table}[t!]
\begin{tabular}{|c|c|l|c|l|}
\hline
No & Price                                                                                           & \multicolumn{1}{c|}{Value}                                                            & \multicolumn{1}{l|}{Noise}                                                                                  & Budget     \\ \hline
C1  & \multirow{4}{*}{\begin{tabular}[c]{@{}c@{}}$i = 3$\\ \\ $j = 4$\\ \\ $\{i, j\} = 7$\end{tabular}} & \begin{tabular}[c]{@{}l@{}}$i=4$, $j=4.9$\\ $\{i,j\}=4.9$\end{tabular}                  & \multirow{4}{*}{\begin{tabular}[c]{@{}c@{}}$i: N(0,1)$\\ \\ \\ \\ $j: N(0,1)$\\ \\\end{tabular}} & Uniform    \\ \cline{1-1} \cline{3-3} \cline{5-5} 
C2  &                                                                                                 & \begin{tabular}[c]{@{}l@{}}$i=4, j=4.1$\\ $\{i,j\}=4.1$\end{tabular}                  &                                                                                                             & Uniform    \\ \cline{1-1} \cline{3-3} \cline{5-5} 
C3  &                                                                                                 & \multirow{2}{*}{\begin{tabular}[c]{@{}l@{}}$i=4, j=4.9$\\ $\{i,j\}=8.7$\end{tabular}} &                                                                                                             & Uniform    \\ \cline{1-1} \cline{5-5} 
C4  &                                                                                                 &                                                                                       &                                                                                                             & Nonuniform \\ \hline
\end{tabular}
\caption{Two item configurations}
\vspace{-2mm}
\label{tab:configs}
\end{table}

\spara{6.2.2 Social welfare}
%\note[Laks]{Some of the algorithms are supposed tp be depricated: e.g., MaxSeqGrd etc. Pls check!}
We now compare the expected social welfare achieved by the algorithms on the four configurations
(Fig. \ref{fig:welfare}). We show the results only for Douban-Movie , since the trend of the results is similar on other networks. In all configuration \sgrd, \sgrdi and \gm outperform all other algorithms. 
The difference in welfare is up to 3$\times$  higher. \mgrd in particular allocates just one of the two items. 
Thus when items exhibit soft competition (C3 and C4), 
%\weic{Why? \mgrd compares all items, so when the difference is high, it will get the higher utility item, and the lost seems to be small. So, are you where C2 is the high difference configuration or the low difference configuration?}
	,\mgrd performs significantly worse. \bc performs comparatively better under soft competition (C3), however for a non-uniform budget again its performance drops. 
\tc on the other hand aims to maximize the adoption count of the item being allocated. Thus it also ends up allocating both the items in same seed nodes. This reduces the overall social welfare for configuration such C1, where both \bc and \tc perform comparatively worse. Social welfare produced by \gm is consistently good, but its running time is exorbitantly high, which prohibits its applicability on any decently sized network. \sgrdi on the other hand is the fastest algorithm, which produces similar welfare across all these four configurations. However, notice that in none of these four configurations, item blocking is effective. We will show later in \textsection 6.3.2 that in the presence of multiple items, when avoiding item blocking is critical, the performance of \sgrdi deteriorates.

\spara{6.2.3 Comparison against \supgrd}
In this experiment we compare \supgrd and \sgrdi on the two largest  networks, \orkut and \twit. We use utility configurations of C1 and C2, but adopt the seed placements needed to meet the constraints required for \supgrd. Recall that for \supgrd the seeds for the inferior items need to be fixed. Hence we select the top 50 nodes using IMM and set them as seeds of $j$. Then, the seeds of $i$ are selected using \supgrd and \sgrdi with the budget being varied from 10 to 50 in steps of 10. We call these new configurations C5 and C6 respectively.

Since the top nodes in terms of the spread are given to $j$, these two cases pose a unique challenge of dealing with arbitrary degree of competition when maximizing welfare. When items' utilities are similar, in C5, new seeds of $i$ should be chosen in a way that minimizes $i$'s overlap with $j$ propagation. Instead in C6, when $i$ has much higher utility, it should be allocated to the top seed nodes. That way, the number of nodes that can be reached by $i$ is much higher and that helps boost the overall social welfare. As can be seen from our results next, that \supgrd can navigate through these varied "strategies", while  \sgrdi cannot.

Fig. \ref{fig:exp_supgrd} (a) and (b) shows the result on the expected social welfare on \orkut and \twit respectively. ``\sgrdi -C5'' (resp. ``\supgrd -C5'') refers to \sgrdi (resp. \supgrd) on C5 and ``\sgrdi -C6'' (resp. ``\supgrd -C6'') on C6. Notice that  in C5 the welfare produced by the two algorithms are comparable. However in C6, where the gap between the individual utilities of the two items is higher, difference between the welfare of \supgrd and \sgrdi is also larger. The reason for that is as follows. \sgrdi uses $\PRIMAP$ to select the seeds of $i$. Consequently to maximize the marginal gain in spread, it minimizes the overlap in the spread of $i$ and $j$ and hence allocates $i$ to lower ranked nodes in terms of spread. However $i$ is the superior item, so allocating  lower ranked nodes to $i$ decreases the overall welfare.

Fig. \ref{fig:exp_supgrd}(c) and (d) compares the running time of the two algorithms on \orkut and \twit. Both the algorithms scale on these large networks. Unlike \sgrdi, running time \supgrd depends on the utility configurations as well. As our running time analysis (\textsection \ref{sec:alg_supgrd}) suggests, when the minimum utility of an item is lower, the running time of \supgrd is higher. However as can be seen, even on large networks, the difference in the running times is not very high: e.g., in configuration C6, the running time of \supgrd is only a 2$\times$  that of \sgrdi, whereas in C5 the running times are similar. To summarize, \supgrd addresses this unique challenge of dealing with an arbitrary degree of competition, with a slightly higher running time.

\subsection{More than two items}

Except for \bc, all the  algorithms can deal with multiple items. In this section, we study their performances when the number of items is more than two. First, we show the impact of increasing the number of items on the running time and social welfare produced. Then we study how the algorithms behave under some challenging configurations designed using multiple items.

\spara{6.3.1 Impact of number of items}
For this experiment, the configuration we test is as follows. Each individual item has expected utility of $1$ and the items exhibit pure competition. Every item has budget $50$ and $\allalloc^p = \emptyset$.
%\weic{In Fig. \ref{fig:assorted}(a) and (b), the x-axis is total budget, so not directly related to the number of items. If you mean that each item has a fixed budget of $50$, perhaps more directly is
%to change the x-axis to be the number of items, and mark it as 1, 2, 3, 4, 5. }

Fig. \ref{fig:assorted}(a) and (b) show respectively, the running time and social welfare produced by the five algorithms. Since \bc cannot run on more than two items, it is omitted. Running time of algorithms \gm, \tc, \mgrd, and \sgrd increases significantly w.r.t the number of items. As the number of items increase, the number of times marginal check is needed for these algorithms, also increases. The marginal check is the most time consuming portion in their running time.
%\weic{Does \tc also use sampling to check marginals? I am not sure. Also, the linear trend or submodular trend is not that obvious, and I feel that we don't have to make such distinction. The important thing is to show the running time difference
%among these algorithms.}
\sgrdi on the other hand relies solely on RR-sets and does not do any marginal checks. Hence the growth in running time is not high. With higher number items, the difference between the running time of \sgrdi and other algorithms, increases.

In terms of social welfare, \tc and \mgrd perform worse than the other algorithms. \mgrd selects only one item in the final allocation, hence it misses out on the additional welfare that could come from allocating the remaining items. Similarly \tc tries to maximize the spread of the last allocated item, at cost of  propagation of other items. Thus their welfare does not increase with more items, unlike the other algorithms.

\begin{table}[]
\begin{tabular}{|l|l|}
\hline
$\util(i) = 2$       & $\util(\{i, j\}) \textless 0$                                                                 \\ \hline
$\util(\{j\}) = 0.11 $    & $\util(\{j, k\}) \textless 0$                                                                 \\ \hline
$\util(\{k\}) = 0.1$     & \multirow{2}{*}{\begin{tabular}[c]{@{}l@{}}$\util(\{i, j, k\})$\\ $\textless{}0$\end{tabular}} \\ \cline{1-1}
$\util(\{i, k\}) = 2.1$ &                                                                                    \\ \hline
\end{tabular}
\vspace{-2mm}
\caption{Three item configuration}
\label{tab:configs_3}
\end{table}

\begin{table}
\begin{tabular}{|r|c|c|c|}
 	\hline
	 item & $p$ & $q$ & $\util_D$ \\ \hline
	$\{indie\}$ 		& $0.107$  & $na$	& $7.0$ \\
	$\{rock\}$			& $0.091$ & $na$	& $6.8$ \\  
	 $\{industrial\}$     & $0.015$ & $na$ & $5.0$\\ 
	 $\{progressive\_metal\}$            &  $0.011$  & $na$ & $4.7$ \\ 
%	 $\{indie, progressive\_metal\}$ &$.0006$ &$-0.0009$ &$1.7$ \\
%	 $\{indie, industrial\}$ & $0.0014$ & $-0.0006$ & $2.6$ \\ 
	 \hline
	\end{tabular} 
	\caption{Learned parameters} \label{tab:real_utils}
\end{table}
\spara{6.3.2 Effect of marginal check}
In our experiments so far, social welfare of \sgrdi has been similar to other algorithms that perform marginal checks. One exception being \supgrd (\textsection 6.2.3), but \supgrd assumes specific constraints that are not general. By not performing the marginal check, \sgrdi runs much faster compared to other algorithms. This begs the question if there is any advantage of using the marginal check altogether. In this experiment we show how marginal check helps avoid item-blocking that \sgrdi fails to circumvent. 

For this experiment, we consider three items in the propagation. Their expected utilities are specified in Table \ref{tab:configs_3}. $i$ has the highest expected utility, followed by $j$ and $k$ has the least. $i$ and $k$ exhibit soft competition hence bundle $\{i, k\}$ has a positive utility, but all other item bundles have negative utilities, exhibiting pure competition. We set the budget of $i$ to 500, and increase the budget of $j$ and $k$ from 100 to 500 each in steps of 100 and study the effect on the welfare produced by \sgrdi and \sgrd.

Fig. \ref{fig:assorted}(c) shows the result on the NetHept network. 
Both algorithms first allocate $i$ as it has the highest individual utility. Then \sgrdi allocates $j$ next, however this allocation is "adjacent" to $i$ \pink{since NetHept is small,} and blocks propagation of $i$ more. Since the utility of $i$ is significantly higher than $j$, allocating $j$ this way in fact causes a negative marginal. \sgrd, using marginal check, postpones allocation of $j$. After $i$, it instead allocates $k$. Although $k$ also has a low individual utility, because of soft competition, it does not block propagation of $i$ and the marginal is non-negative. It later allocates $j$, which is now further apart from $i$, hence cannot block $i$'s propagation. Thus \sgrd produces a social welfare which is higher than that of \sgrdi. Further, as the budget of $j$ increases, the amount of blocking also increases, hence the welfare difference between the two algorithms also goes up.

\spara{6.3.3 Scalability of \sgrdi}
Our next experiment shows the impact of network size on $\sgrdi$ using $\orkut$ with two types of edge probabilities: (1) $1/\ind(v)$ and (2) fixed $0.01$.  
We use a uniform budget of $50$ for all three items. Instead of using the full network, we use breadth-first-search to progressively increase the network size so that it includes a certain percentage of the total nodes in the network. At $100\%$, the full network is used. Fig. \ref{fig:assorted}(d) shows the results. ``\sgrdi, time 1'' and ``\sgrdi, time 2'' depict the running time of \sgrdi on the two types of edge probabilites respectively. As the network size increases, the running time in both cases roughly has a linear increase.

%\note[Laks]{Any reason why supgrd and seqgrd-nm are not compared for time and welfare on the same dataset orkut?}

%\note[Laks]{When first introducing the inverse in-degree assignment earlier, should we use that name, which is standard in the literature?}

%\note[Laks]{Fig. and Figure --  Make it consistent!}

\subsection{Real item experiments}

In this section, we learn the utilities of items from real dataset instead of the synthetic utilities used in earlier experiments. 
The dataset used is the $\lfg$ generated from the listening behavior of users of the music streaming service Last.fm  \cite{lastfm1k-data,lastfmtags-data}. 
This dataset was used in \cite{benson2018discrete} to learn the adoption probabilities of different items, where each genre is treated as an item. 
This dataset also echos our first motivating example in the introduction.
%	it is the only dataset where a significant number of partial-competition is observed. 
%\weic{Do we really have partial copetition? Table 4 suggests that all are pure competition? If the data we learned are only pure competition,
%	perhaps we should not mention this.}
%In contrast the other datasets mostly exhibit pure competion as the only mode of competition (Fig. 2 in \cite{benson2018discrete}). 
We next establish the connection between the adoption probabilities and the utilities, which enables us to learn the parameters using \cite{benson2018discrete}.

\spara{6.4.1 Learning the utilities}
In \cite{benson2018discrete}, every item $i$ is associated with an adoption probability $p_i$. Adoption probability of an itemset $I = \{i, ..., k\}$ is $p_{I} = \gamma_{|I|} \prod_{j \in I}p_{j} + q_{I}$, where $q_{I}$ is a correction received depending on the way  items in  $I$ interact with each other: if the items are complementary, then the correction is positive, if competing then it is negative, and $0$ if the items are independent. These probabilities and corrections are learnt in \cite{benson2018discrete} from the dataset of how frequently items are selected together by the users. 

According to Observation 2.2 of \cite{benson2018discrete}, $ p_i = e^{v_i} / \sum_j e^{v_j} $, where $v_i$ is the expected utility of item $i$ as per our utility model. 
Given a set of learnt $p_i$, we first set $\sum_j e^{v_j} = 10000$. Then for every $i$, we set $v_i = \ln(10000 \cdot p_i)$. We choose the number $10000$ to ensure that the corresponding utilities are positive. Finally we set the expected utility of item $i$, $\util(i) = v_i$.
Next for an itemset $I = \{i, ..., k\}$, \cite{benson2018discrete} learns two parameters $\gamma_{|I|}$ and $q'_{I}$. By using $q_{I} = \gamma \cdot q_{I}'$ the probability of adopting the bundle, $p_{I}$ is derived. The expected utility of the bundle is similarly set to be $\util(I) = \ln(10000 \cdot p_{I})$. Notice that the exact values of utilities are not as important as the relative order of utilities of different itemsets. The way utilities are learnt is in correspondence with the adoption probabilities learned in [7]. 

Table \ref{tab:real_utils} shows the utilities of four different items (i.e., genres) in the dataset: \textit{rock, indie, industrial and progressive\_metal},  learned using the above described method. 
Larger bundles are either not present in the dataset or have smaller learned utilities compared to the individual items in the bundle,
	suggesting that items are in pure competition in our utility model.

\eat{
\begin{table}[t!]
	\scriptsize

\subfloat[Three item configuration] \label{tab:configs_3}
{	\begin{tabular}{|l|l|}
\hline
$\util(i) = 2$       & $\util(\{i, j\}) \textless 0$                                                                 \\ \hline
$\util(\{j\}) = 0.11 $    & $\util(\{j, k\}) \textless 0$                                                                 \\ \hline
$\util(\{k\}) = 0.1$     & \multirow{2}{*}{\begin{tabular}[c]{@{}l@{}}$\util(\{i, j, k\})$\\ $\textless{}0$\end{tabular}} \\ \cline{1-1}
$\util(\{i, k\}) = 2.1$ &                                                                                    \\ \hline
\end{tabular}
%\vspace{-2mm}
}
\subfloat[Learnt parameters] \label{tab:real_utils}{
	\begin{tabular}{r|c|c|c|}
	 item & $p$ & $q$ & $\util_D$ \\ \hline
	$\{indie\} (i)$ 		& $0.107$  & $na$	& $7.0$ \\
	$\{rock\} (j)$			& $0.091$ & $na$	& $6.8$ \\  
	 $\{industrial\} (k)$     & $0.015$ & $na$ & $5.0$\\ 
	 $\{progressive\_metal\} (l)$            &  $0.011$  & $na$ & $4.7$ \\ 
	 $\{indie, progressive\_metal\}$ &$.0006$ &$-0.0009$ &$1.7$ \\
	 $\{indie, industrial\}$ & $0.0014$ & $-0.0006$ & $2.6$ \\ \hline
	\end{tabular}
	}
%\end{table}
}

\captionsetup[table]{font=normalsize}
\begin{table*}[h]
\centering
\resizebox{\linewidth}{!}{%
\begin{tabular}{|l|l|l|l|l|l|l|l|l|l|l|l|} 
\hline
\multirow{2}{*}{Network} & \multirow{2}{*}{Budget} & \multirow{2}{*}{Algorithm} & \multicolumn{5}{l|}{Real Utility Configuration (as shown in Table~\ref{tab:real_utils})} & \multicolumn{4}{l|}{Synthetic Utility Configuration (as shown in Table~\ref{tab:configs_3})} \\ 
\cline{4-12}
 &  &  & indie & rock & industrial & progressive\_metal & welfare & i & j & k & welfare \\ 
\hline
\multirow{3}{*}{NetHEPT} & \multirow{3}{*}{10} & RR & 203 & 217 & 191 & 196 & 4473.08 & 277 & 244 & 234 & 513.2 \\ 
\cline{3-12}
 &  & Snake & 204(+0.005) & 201(-0.081) & 207(+0.083) & 195(-0.005) & 4458.64(-0.003) & 258(-0.068) & 246(+0.008) & 261(+0.115) & 478.4(-0.068) \\ 
\cline{3-12}
 &  & SGRD-NM & 255(+0.252) & 199(-0.082) & 188(-0.016) & {\color{green}\textbf{165(-0.158)}} & 4951.8 (+0.112) & 306(+0.105) & 220(-0.098) & 227(-0.030) & 577.4(+0.125) \\ 
\hline
\multirow{3}{*}{NetHEPT} & \multirow{3}{*}{40} & RR & 496 & 493 & 491 & 475 & 10795.3 & 667 & 576 & 645 & 1227.3 \\ 
\cline{3-12}
 &  & Snake & 483(-0.026) & 496(+0.006) & 488(-0.004) & 488(+0.027) & 10758.2(-0.003) & 648(-0.028) & 581(+0.009) & 669(+0.037) & 1194.5(-0.027) \\ 
\cline{3-12}
 &  & SGRD-NM & 673(+0.357) & 499(+0.012) & 419(-0.147) & {\color{green}\textbf{365(-0.189)}} & 11264.5(+0.043) & 800(+0.199) & 510(-0.114) & 514(-0.203) & 1510.6(+0.230) \\ 
\hline
\multirow{3}{*}{Orkut} & \multirow{3}{*}{10} & RR & 37790 & 38888 & 38331 & 34711 & 828368.2 & 69151 & 49730 & 67405 & 110032.5 \\ 
\cline{3-12}
 &  & Snake & 38241(+0.012) & 37401(-0.038) & 39818(+0.039) & 34260(-0.013) & 828235.4(-0.002) & 67648(-0.021) & 50511(+0.016) & 68510(+0.016) & 107227.7(-0.026) \\ 
\cline{3-12}
 &  & SGRD-NM & 50800(+0.344) & 40837(+0.050) & 31189(-0.186) & 26895(-0.225) & 864154.3(+0.040) & 76784(+0.110) & 50219(+0.010) & 57199(-0.151) & 124210.9(+0.129) \\ 
\hline
\multirow{3}{*}{Orkut} & \multirow{3}{*}{40} & RR & 58142 & 58586 & 59939 & 54607 & 1276650.6 & 119039 & 83291 & 113359 & 183853.2 \\ 
\cline{3-12}
 &  & Snake & 57211(-0.016) & 56922(-0.028) & 61603(+0.028) & 55538(+0.017) & 1272190.7(-0.035) & 117454(-0.013) & 82937(-0.043) & 115338(+0.018) & 180269.4(-0.020) \\ 
\cline{3-12}
 &  & SGRD-NM & 106876(+0.838) & 54909(-0.063) & 42218(-0.296) & {\color{green}\textbf{27272(-0.501)}} & 1397770.8(+0.095) & 150926(+0.268) & {\color{green}\textbf{63577(-0.237)}} & 87480(-0.228) & {\color{green}\textbf{253427.9(+0.378)}} \\
\hline
\end{tabular}
}
\caption{Adoption count of different items and the overall social welfare}\label{tab:utilvsadopt}
\end{table*}
\spara{6.4.2 Results using real parameters}

We use the learned utility configuration to compare the social welfare produced by the algorithms on two networks, namely \net and \orkut. For the experiment we set uniform budget for all the four items, which varies from 10 to 40 in steps of 10. The algorithms compared are $\tc, \mgrd, \sgrd$ and $\sgrdi$. The results are shown in Fig \ref{fig:real_exp}.

In terms of running time the results are similar to our previous experiments (Fig \ref{fig:real_exp}(a)-(b)). $\sgrdi$ outperforms the other algorithms by orders of magnitude, since it does not require the time consuming marginal gain computation. For social welfare, notice that the real utility configuration exhibits pure competition. 
As noted earlier, under pure competition, social welfare produced by $\sgrd$ and $\sgrdi$ coincide. $\mgrd$ and $\tc$ on the other hand typically encourage the adoption of one single item. 
Hence the difference in social welfare produced by these algorithms compared to $\sgrdi$ is higher since the number of items are also more than the previous configurations we used.

\spara{6.4.3 Social welfare vs adoption}

Our final set of experiments compare the relationship between the social welfare and item adoptions. In particular, we want to investigate whether maximizing welfare for competing items could result in a significant drop in the number of item adoptions. For this experiment, we focus on two utility configurations -- (i) Real utility of Table \ref{tab:real_utils}, which exhibits pure competition and (ii) Synthetic utility of Table \ref{tab:configs_3}, which exhibits a mix of partial and pure competition. NetHEPT and \orkut are the two networks used and each item's budget is set to two different values, 10 and 40. 

We compare our algorithm $\sgrdi$ against two baselines. After selecting the seed nodes, the first baseline allocates items to the nodes in a round robin manner, hence it is called $\rr$. 
The second baseline, called $\sn$, is similar to \rr, but it flips the order for every successive sequence of allocations. 
To illustrate, if there are 4 seed nodes $s_1, ..., s_4$, in order, and two items $i, j$, $\sgrdi$ allocates as $s_1\colon i,s_2\colon i,s_3\colon j,s_4\colon j$, {\rr} allocates as $s_1\colon i,s_2\colon j,s_3\colon i,s_4\colon j$ and {\sn} allocates as $s_1\colon i,s_2\colon j,s_3\colon j,s_4\colon i$. 
Table \ref{tab:utilvsadopt} shows the adoption count of each item and the social welfare produced by these algorithms under different configurations.

In terms of the social welfare objective, $\sgrdi$ dominates across all different configurations. {\rr} produces the next highest welfare. Hence we report the fractional change ($+$ denotes increase and $-$ denotes decrease), in comparison to {\rr}, next to each entry of the table. The entries that deserve more attention are highlighted in green.

As can be seen, the total number of adoptions of all the items remains the same across all three algorithms. However, \sgrdi generally increases the adoption of the superior product to increase the welfare, while reducing the adoption of the inferior item. On NetHEPT, for budget 10, the maximum drop in adoptions happens for the most inferior item (progressive\_metal), by  $15.8\%$. For a higher budget, the drop increases (to $18.9\%$), because when budget increases for the superior item, $\sgrdi$ allocates lower ranked seeds for the inferior item.

The highest drop in adoption i.e., $50.1\%$, also happens for the item progressive\_metal for budget 40 on Orkut. This is because when items are purely competing, number of items is high and each item has a large budget, the inferior items' seeds are in fact much lower ranked. However, if it exhibits partial competition with a superior item, then leveraging it the adoption does not decrease that much. That is why in Orkut even when the budget is 40, for the synthetic utility configuration, the highest drop in adoptions for the inferior items is only $23.7\%$. Also notice $\sgrdi$ produces significantly higher social welfare compared to the baselines, the increase being up to $37.8\%$.  
In summary, we see that our welfare maximization algorithm  provide more adoptions to the superior items and fewer adoptions to the inferior
	items, but the amount of change is not too drastic. We argue that this is the "price" of enhancing  the overall user satisfaction; also the drop in the adoptions of the inferior items is exactly because they are not as competitive.

%It is worth noticing that even for the entire million-sized network and fixed probability, $\sgrdi$ requires mere $259$ seconds to complete, which again attests to its scalability. 

%\note[Laks]{Shd we tone this down? The \#seeds we consider is small. We can def. say linear scalability in network size.} 

%\note[Laks]{What is \sgrdi time 1 and time 2 in the plots? Not explained anywhere.}

%\weic{There is no overall conclusion in the experiment section. That is, which one we should promote after seeing our experimental results?
%	I see some similar conclusion in the next section, but think it may be better to state it here.
%So I added the paragraph below, and simplify the conclusion section.}

To conclude this section, we generally observe that:
(a) when the conditions required by \supgrd are met, it is the best option providing the best social welfare and competitive running time;
(b) in the general case, \sgrdi performs well in most cases and has the best running time, but when item blocking is significant, its marginal-checking version \sgrd 
	could provide better social welfare, at the cost of higher running time; 
(c) \mgrd could be used to enhance the theoretical guarantee when the utility difference is high, but its superiority is not typically observed in large networks; 
(d) our algorithms outperform all baselines on social welfare and running time and scale to large networks. Our algorithms achieve superior welfare at the expense of a reasonable drop in the adoption count of inferior items, keeping the total adoption count unchanged. 

\vspace{-3mm}
\section{Conclusions and future work} \label{sec:concl} 

In this paper, we study the problem of maximizing social welfare over competing items under the UIC  model. The problem is not only NP-hard but  is also NP-hard to approximate within a constant factor. Further we find that due to conflicting requirements, it is challenging to design a single algorithm that can work effectively for all different utility configurations. 
Yet we propose a cohort of efficient algorithms that not only provide approximation guarantees but also scale well to real large networks, and their performance  is validated through extensive experiments on real-world networks.
%
%\sgrdi in particular is the most general algorithm that is shown to have good empirical performance both in terms of welfare and running time. It also handles multi-item propagation's effectively as well. When only inferior items' seeds are fixed \supgrd perform better in terms of the welfare while incurring a slight increase in the running time, however it still scales for all the large networks. Thus, between these two algorithms almost all the major practical cases of competitions are covered.

Although  welfare maximization under competition ensures that users' total  utility from adoptions is maximized, it does not directly ensure fairness. For a campaigner who often pays for advertising, ensuring that her item is seen at least by a certain number of users is critical. While fairness in IM  has been studied recently,   incorporating fairness in social welfare maximization will be an interesting  challenge. Further, this paper and \cite{ban2019} studied competition and complementarity in isolation. Designing algorithms for an arbitrary mix of competing and complementary items is an intriguing  problem.

\eat{ 
\vspace*{-1ex} 
\section{Summary \& Discussion}\label{sec:concl}
\input{sec-concl}
} 

\clearpage  
{
\bibliographystyle{abbrv}
\bibliography{kdd2020-cepic}  % sigproc.bib is the name of the Bibliography in this case
}

\end{document}